\documentclass[12pt, a4paper]{article}
\usepackage[T1]{fontenc}

\usepackage[a4paper, left=1in, top=1in, right=1in, bottom=.8in, nohead, includefoot]{geometry}
\usepackage{appendix, charter, graphicx, natbib, setspace, multirow, amsmath, amssymb, amsfonts, enumitem}

\usepackage[dvipsnames]{xcolor}
\usepackage{bm}
\usepackage{graphicx}
\usepackage{diagbox}
\usepackage{array}
\usepackage{setspace}
\usepackage{rotating}
\usepackage{multirow}
\usepackage{booktabs}
\usepackage{color}
\usepackage{setspace}
\usepackage{enumitem}
\usepackage[colorlinks=true, urlcolor=blue, linkcolor=blue, citecolor=blue]{hyperref}
\usepackage{rotating}

\usepackage{sectsty}
\allsectionsfont{\normalsize}

\makeatletter

\@addtoreset{equation}{section}
\makeatother

\usepackage{caption}
\usepackage[labelformat=simple]{subcaption}
\makeatletter\let\p@subfigure\thefigure\makeatother

\usepackage{cleveref}
\crefname{chapter}{Chapter}{Chapters}
\crefname{section}{Section}{Sections}
\crefname{subsection}{Section}{Sections}
\crefname{subsubsection}{Section}{Sections}
\crefname{figure}{Figure}{Figures}
\crefname{table}{Table}{Tables}
\crefname{equation}{Equation}{Equations}
\crefname{appendix}{Appendix}{Appendices}

\def\equationautorefname~#1\null{
Eq.~(#1)\null
}

\newcolumntype{d}[1]{D{.}{.}{#1}}

\title{\textbf{\LARGE Should I stay or should I go? \\[.3em] A latent threshold approach to large-scale\\[-.3em] mixture innovation models\footnote{This work is a substantially revised version of a paper circulated under the title ``Should I stay or should I go? Bayesian inference in the threshold time-varying parameter model.''
}%
}}
\usepackage{authblk}

\author[1]{Florian Huber}
\author[1]{Gregor Kastner\footnote{Corresponding author. Address: Institute for Statistics and Mathematics, WU Vienna University of Economics and Business, Welthandelsplatz 1, Building D4, 4th Floor, 1020 Wien, Austria. Phone: +43 1 31336-5593. Fax: +43 1 31336-905593. Email: \href{mailto:gregor.kastner@wu.ac.at}{gregor.kastner@wu.ac.at}.}}
\author[2]{Martin Feldkircher\thanks{The opinions expressed in this paper are those of the authors and do not necessarily reflect the official viewpoint of the Oesterreichische Nationalbank or the Eurosystem.}}
\affil[1]{WU Vienna University of Economics and Business}
\affil[2]{Oesterreichische Nationalbank (OeNB)}



\begin{document}
\graphicspath{{Figs/}}
\maketitle

\begin{abstract}
\noindent
This paper proposes a straightforward  algorithm to carry out inference in large time-varying parameter vector autoregressions (TVP-VARs) with  mixture innovation components for each coefficient in the system. We significantly decrease the computational burden by approximating the latent indicators that drive the time-variation in the coefficients with a latent threshold process that depends on the absolute size of the shocks. The merits of our approach are illustrated with two applications. First,  we forecast the US term structure of interest rates and demonstrate forecast gains of the proposed mixture innovation model relative to other benchmark models.  Second, we apply our approach to US macroeconomic data and find significant evidence for time-varying effects of a monetary policy tightening. 

\end{abstract}






\newpage

\section{Introduction}
\label{sec:intro}
In the last few years, economists in policy institutions and central banks were criticized for their failure to foresee the recent financial crisis that engulfed the world economy and led to a sharp drop in economic activity. Critics argued that economists failed to predict the crisis because  models commonly utilized at policy institutions back then were too simplistic. For instance, the majority of forecasting models adopted were (and possibly still are) linear and low dimensional. The former implies that the underlying structural mechanisms and the volatility of economic shocks are assumed to remain constant over time -- a rather restrictive assumption. The latter implies that  only little information is exploited which may be detrimental for obtaining reliable predictions.

In light of this criticism,  practitioners started to adopt more complex models that are capable of capturing salient features of time series commonly observed in macroeconomics and finance. These models are based on earlier research that provides considerable evidence, at least for US data, that the influence of certain variables appears to be time-varying  \citep{stock1996evidence,cogley2002evolving,cogley2005drifts,primiceri2005time,sims2006were}. 
This raises additional issues related to model specification and estimation. For instance, do all regression parameters vary over time? Or is time variation just limited to a specific subset of the parameter space? Moreover, as is the case with virtually any  modeling problem, the question  whether a given variable should be included in the model in the first place  naturally arises. Apart from deciding whether parameters are changing over time, the nature of the process that drives the dynamics of the coefficients also proves to be an important modeling decision.

In a recent contribution, \cite{fruhwirth2010stochastic} focus on model specification issues within the general framework of state space models. Exploiting a non-centered parametrization of the model allows them to rewrite the model in terms of a constant parameter specification, effectively capturing the steady state of the process along with deviations thereof. The non-centered parameterization is subsequently used to search for appropriate model specifications, imposing shrinkage on the steady state part and the corresponding deviations. 

Recent research aims to discriminate between inclusion/exclusion of elements of different variables and whether the associated regression coefficients are constant or time-varying \citep{koop2012forecasting,koop2013large, kalli2014time,belmonte2014hierarchical, eisenstat2016stochastic}.  Another strand of the literature asks whether coefficients are constant or time-varying by assuming that the innovation variance in the state equation is characterized by a change point process\citep{mcculloch1993bayesian,gerlach2000efficient, koop2009evolution, giordani2012efficient}. However, the main drawback of this modeling approach is the severe computational burden originating from the need to simulate additional latent states for each parameter. This renders estimation of large dimensional models like vector autoregressions (VARs) unfeasible. To circumvent such problems, \cite{koop2009evolution} estimate a single Bernoulli random variable to discriminate between time constancy and parameter variation for the autoregressive coefficients, the covariances, and the log-volatilities, respectively. This assumption, however, implies that either all autoregressive parameters change over a given time frame, or none of them. Along these lines, \cite{mah-son:eff} allow for independent breaks in regression coefficients and the volatility parameters. However, they show that their multivariate approach is inferior to univariate change point models when out-of-sample forecasts are considered and conclude that allowing for independent breaks in each series is important.

In the present paper, we introduce a method that can be applied to a highly parameterized  VAR model by combining ideas  from the literature on  latent threshold models \citep{nee-dun:bay, nakajima2013bayesian,nakajima2013dynamic,zhou2014bayesian,kimura2016identifying} to approximate the latent indicators during Markov chain Monte Carlo (MCMC) sampling. As mentioned above, the main computational hurdle stems from the necessity to apply forward-filtering backward-sampling (FFBS) based algorithms to estimate the indicators that control the time-variation in the regression coefficients. 

The key contribution of this paper is to avoid computationally intensive simulation of the latent indicators by proposing a straightforward approximation to these indicators and thus allow for estimation of large-scale models.  In doing so, we mimic the behavior of a standard mixture innovation model by setting the value of an indicator equal to one if the absolute value of the parameter change exceeds a threshold to be estimated. In that case, the corresponding state innovation variance is set to a large value, allowing for large jumps in the regression coefficients. By contrast, if the absolute changes are small (i.e., below the threshold), a state innovation variance close to zero is adopted and the corresponding regression coefficient can be viewed as being constant over that certain stretch in time. Compared to existing algorithms, the additional costs of estimating the proposed model, henceforth labeled the threshold time varying parameter (TTVP) model, is negligible. To assess systematically, in a data-driven fashion, which predictors should be included in the model, we impose a set of Normal-Gamma priors \citep{griffin2010inference} in the spirit of \cite{bitto2015achieving} on the initial state of the system. The TTVP code is bundled in the R
package \texttt{threshtvp}, which is made available from the authors upon request.

We illustrate the empirical merits of our approach by carrying out two empirical exercises. In the first exercise, we predict the US term structure of interest rates.  The proposed framework is benchmarked against several constant parameter Bayesian VAR models with stochastic volatility (SV) and hierarchical shrinkage priors, time-varying parameter VARs as well as a multivariate random walk with SV. Moreover, we follow \cite{diebold2006forecasting} and use a model based on the \cite{nelson1987parsimonious} three factor model.  The findings indicate that our proposed TTVP specification outperforms all competing specifications for one-month-ahead as well as three-month-ahead predictions.  The forecasting gains appear to be especially pronounced during crisis episodes. 

In the second  application,  we use a medium-scale US macroeconomic dataset to investigate the degree of time-variation  of the underlying causal mechanisms for the US.  Considering the determinant of the time-varying variance-covariance matrix of the state innovations as a global measure for the strength of parameter movements, we find that these movements reach a maximum in the beginning of the 1980s. This is driven by effects on inflation for which we find a considerable price puzzle in the 1960s which starts disappearing in the early 1980s. Effects on other variables such as output and investment growth as well as hours worked vary more gradually over time. These effects are especially pronounced during the aftermath of the global financial crisis in 2008/09 indicating  evidence for increased effectiveness of monetary policy. 

The paper is structured as follows.
Section~\ref{sec:framework} introduces the  modeling approach, the prior setup  and the corresponding MCMC algorithm for posterior simulation. Section~\ref{sec:illustration} illustrates the behavior of the model by showcasing scenarios with no, few, and many jumps in the state equation, alongside a standard TVP specification with sustained movement. In Section~\ref{sec:application1}, we predict the US term structure of interest rates. In Section~\ref{sec:application2}, we apply the model to a medium-scale US macroeconomic dataset and  investigate during which periods VAR coefficients display the largest amount of time-variation; furthermore, we scrutinize the associated implications on dynamic responses with respect to a monetary policy shock. Finally, Section~\ref{sec:conclusion} concludes.
\section{Econometric framework}
\label{sec:framework}
In this section, we first introduce a univariate dynamic regression model  that is capable of discriminating between constant and time-varying parameters at each point in time.  This stylized framework is used to discuss the main ideas of the paper. We then subsequently generalize this model framework to the VAR case that is used in the empirical applications.
\subsection{A mixture innovation model}
\label{sec:model}
Consider the following dynamic regression model,
\begin{equation}
y_t = \boldsymbol{x}_t' \boldsymbol{\beta}_t + u_t, \quad u_t \sim \mathcal{N}(0, \sigma_t^2)\label{eq:obs},
\end{equation}
where $\boldsymbol{x}_t$ is a $K$-dimensional vector of explanatory variables and $\boldsymbol{\beta}_t =(\beta_{1t},\dots,\beta_{Kt})'$ a vector of regression coefficients. The error term $u_t$ is assumed to be independently normally distributed with (potentially) time-varying variance.
This model assumes that the relationship between elements of $\boldsymbol{x}_t$ and $y_t$ is not necessarily constant over time, but changes subject to some law of motion for $\boldsymbol{\beta}_t$. Typically, researchers assume that the $j$th element of $\boldsymbol{\beta}_t, \beta_{jt}~(j = 1,\dots,K)$, follows a random walk process,
\begin{equation}
\beta_{jt} = \beta_{j,t-1}+e_{jt}, \quad e_{jt} \sim \mathcal{N}(0,\vartheta_j), \label{eq:states1}
\end{equation}
with $\vartheta_j$ denoting the innovation variance of the latent states. Equation (\ref{eq:states1}) implies that parameters evolve gradually over time, ruling out abrupt changes. While being conceptually flexible, in the presence of only a few breaks in the parameters, this model generates spurious movements in the coefficients that could be detrimental for the empirical performance of the model \citep{d2013macroeconomic}.

Thus, we deviate from \autoref{eq:states1} by specifying the innovations of the state equation $e_{jt}$ to be a mixture distribution. More concretely, let
\begin{align}
e_{jt} &\sim \mathcal{N}(0, \theta_{jt}), \label{eq:threshold_1} \\
\theta_{jt} &=s_{jt} \vartheta_{j1}+(1-s_{jt}) \vartheta_{j0}, \label{eq:threshold_1b}
\end{align}
where $\vartheta_{j1}$ and $\vartheta_{j0}$ are state innovation variances with $\vartheta_{j1} \gg \vartheta_{j0}$ and $\vartheta_{j0}$ set close to zero. Furthermore,  $s_{jt}$ is an indicator variable that follows a Bernoulli distribution, i.e.,
\begin{equation}
s_{jt} = \begin{cases} 1~\text{ with probability }~p_j, \\
0~\text{ with probability }~1-p_j. \end{cases}
\end{equation}
This model
is a relatively standard mixture innovation model \citep{mcculloch1993bayesian, gerlach2000efficient,giordani2012efficient}.\footnote{The main difference is that the literature typically assumes that $\vartheta_{j0}\equiv0$ for all $j$ \citep[for an exception, see][]{carter1994gibbs}.} \autoref{eq:threshold_1} states that  if $s_{jt}$ equals one, we assume that the change in $\beta_{jt}$ is normally distributed with  zero mean and variance $\vartheta_{j1}$.  On the contrary, if $s_{jt}$ equals zero, the innovation variance is set close to zero, effectively implying that $\beta_{jt} \approx \beta_{j,t-1}$, i.e.,~almost no change from period $(t-1)$ to $t$.  

This modeling approach provides a great deal of flexibility, nesting a plethora of simpler model specifications. The interesting cases are characterized by situations where  $s_{jt} = 1$ only for some $t$. For instance, it could be the case that parameters tend to exhibit strong movements at given points in time but stay constant for the majority of the time. An unrestricted time-varying parameter model would imply that the parameters are gradually changing over time, depending on the innovation variance in \autoref{eq:states1}. Another prominent case would be a structural break model with an unknown number of  breaks \citep[for a Bayesian exposition, see e.g.][]{koop2007estimation}. Recently, this framework has been extended by \cite{uri-lop:dyn} who model the indicator as a first-order two-state Markov process.

\subsection{Mitigating the computational burden through thresholding}\label{sec: approx}
Unfortunately, estimation of the model described in the previous section is computationally cumbersome if $K$ is large as in  multivariate systems like VARs, even though there exist several estimation strategies. One strand of the literature \citep[see][]{mcculloch1993bayesian} estimates the indicators conditional on the states using  single-step Gibbs updating within a larger MCMC algorithm. This, however, often results in poor mixing properties of the algorithm since the states and the indicators are typically highly correlated. The more recent literature \citep{gerlach2000efficient} simulates the indicators after integrating out the latent states using Kalman-filter-based algorithms.   Unfortunately, this procedure has to be repeated  for each coefficient during MCMC sampling, turning computationally prohibitive even for moderate $K$. Thus, researchers often resort to models where only a small number of  indicators is introduced that determines the amount of time-variation for certain parts of the parameter space in the system \citep[for a VAR application, see,][]{koop2009evolution}.

The key innovation of the present paper is to circumvent this issue by proposing a relatively simple approximation that makes immediate use of the fact that Gibbs sampling generates draws from the joint posterior by sampling from the full conditionals. Similarly to the early literature on mixture innovation models mentioned above \citep{mcculloch1993bayesian}, we also condition on the states to simulate the indicators $s_{jt}$ during MCMC simulation.
However, instead of directly sampling from this full conditional distribution, we introduce one additional auxiliary parameter per coefficient, the threshold $d_j$, which in turn renders the indicators conditionally deterministic. More concretely, in the $l$th iteration of our MCMC algorithm, after obtaining draws $\{\beta_{jt}^{(l)}\}_{t=1,\dots,T}$ conditional on draws of the indicators $\{s_{jt}^{(l-1)}\}_{t=1,\dots,T}$ and the remaining parameters, we approximate $s_{jt}$ through
\begin{equation}
\hat{s}^{(l)}_{jt} = \begin{cases} 1 ~\text{ if }~ |\Delta \beta^{(l)}_{jt}|>d^{(l-1)}_j, \\
0 ~\text{ if } ~ |\Delta \beta^{(l)}_{jt}|\le d_j^{(l-1)}, \end{cases} \label{eq:threshold_2}
\end{equation}
where $d^{(l-1)}_j$ denotes the $(l-1)$th draw of a coefficient-specific threshold $d_j$ to be estimated and $\Delta \beta^{(l)}_{jt} := \beta^{(l)}_{jt}-\beta^{(l)}_{j,t-1}$.  \autoref{eq:threshold_2} states that  if the absolute period-on-period change of  the $l$th draw of $\beta_{jt}$ exceeds the $(l-1)$th draw of the threshold $d_j$, we set $\hat{s}^{(l)}_{jt}=1$ and thus use a large variance. By contrast, if the change in the current draws of the parameter is too small,  the innovation variance is set close to zero, effectively implying that $\beta_{jt} \approx \beta_{j,t-1}$.
The detailed description of the MCMC sampler, along all required full conditionals, can be found in Section~\ref{sec:mcmc}.



Compared to a standard mixture innovation model that postulates $s_{jt}$ as a sequence of independent Bernoulli variables, our approach, labeled the threshold mixture innovation model, mimics this behavior by assuming that regime shifts are governed by a deterministic law of motion, conditionally on the current draw of $\{\beta_{jt}\}_{t=1,\dots,T}$ and $d_j$. The main advantage of our approach relative to standard mixture innovation models is that instead of having to estimate a full sequence of $s_{jt}$ for all $j$, the proposed framework only relies on a single additional parameter per coefficient. This renders estimation of high dimensional models such as vector autoregressions (VARs) feasible. The additional computational burden turns out to be negligible relative to an unrestricted TVP-VAR, see again Section~\ref{sec:mcmc} for more information.

Our model is also related to the latent thresholding approach put forward in  \cite{nakajima2013bayesian} within the time series context. However, while in their model latent thresholding discriminates between the inclusion or exclusion of a given covariate at time $t$, our model uses information on the changes in a given regression coefficient to mimic the behavior of a mixture innovation model. In addition, while the model proposed in \cite{nakajima2013bayesian} assumes that the thresholded process enters \autoref{eq:obs} directly, our approach is based on estimating a non-linear model for the state equation. Nevertheless, notice that if the indicators are treated as augmented data, our model is a conditionally linear Gaussian state space model and thus standard  algorithms can be used to estimate the latent states.

\subsection{A multivariate extension with stochastic volatility} \label{sec: VAR_part}
The model proposed in the previous subsection can be straightforwardly generalized to the VAR case with multivariate SV by letting $\boldsymbol{y}_t$ be an  $m$-dimensional response vector. In this case, \autoref{eq:obs} becomes
\begin{equation}
\boldsymbol{y}_t = \boldsymbol{x}_t' \boldsymbol{\beta}_t+\boldsymbol{u}_t,
\end{equation}
with $\boldsymbol{x}'_t = \{\boldsymbol{I}_m \otimes \boldsymbol{z}'_t\}$, where $\boldsymbol{z}_t = (\boldsymbol{y}'_{t-1}, \dots, \boldsymbol{y}'_{t-P})'$ includes the $P$ lags of the  endogenous variables.\footnote{In the empirical application, we also include an intercept term which we omit here for simplicity.} The vector $\boldsymbol{\beta}_t$ now contains the dynamic autoregressive coefficients with dimension $K=m^2P$ where each element follows  the state evolution given by Eqs.~(\ref{eq:states1}) to (\ref{eq:threshold_2}). The vector of white noise shocks $\boldsymbol{u}_t$ is distributed as
\begin{equation}
\boldsymbol{u}_t \sim \mathcal{N}(\boldsymbol{0}_m, \boldsymbol{\Sigma}_t).
\end{equation}
Hereby, $\boldsymbol{0}_m$ denotes an $m$-variate zero vector and $\boldsymbol{\Sigma}_t = \boldsymbol{V}_t \boldsymbol{H}_t \boldsymbol{V}_t'$ is a time-varying variance-covariance matrix. The matrix $\boldsymbol{V}_t$ is a lower triangular matrix with unit diagonal and $\boldsymbol{H}_t = \text{diag}(e^{h_{1t}},\dots,e^{h_{mt}})$. We assume that the logarithm of the variances evolves according to
\begin{equation}
h_{it} = \mu_i + \rho_i (h_{i,t-1}+\mu_i) + \nu_{it}, \quad i=1,\dots,m, \label{eq: logvolas}
\end{equation}
where $\mu_i$ and $\rho_i$  are equation-specific mean and persistence parameters and $\nu_{it} \sim \mathcal{N}(0,\zeta_i)$ is  an equation-specific white noise error with variance $\zeta_i$. For the covariances in $\boldsymbol{V}_t$ we impose random walk state equation with thresholded error variances in analogy to \autoref{eq:threshold_1}.

Conditional on the ordering of the variables, it is straightforward to estimate the  model on an equation-by-equation basis, augmenting the $i$th equation with the contemporaneous values of the preceding $(i-1)$ equations, leading to a Cholesky-type decomposition of the variance-covariance matrix. Thus, the  $i$th equation (for $i=2,\dots,m$) is given by
\begin{equation}
y_{it} =  \tilde{\boldsymbol{z}}'_{it} \tilde{\boldsymbol{\beta}}_{it}+  u_{it}. \label{eq: vareqspecific}
\end{equation}
Here, $\tilde{\boldsymbol{z}}_{it}= (\boldsymbol{z}'_t, y_{1t},\dots, y_{i-1, t})'$ denotes the augmented vector of regressors, while $\tilde{\boldsymbol{\beta}}_{it}=(\boldsymbol{\beta}_{it}', \tilde{v}_{i1, t}, \dots, \tilde{v}_{i, i-1, t})'$ is a vector of latent states with dimension $K_i= mP+i-1$ where $\boldsymbol{\beta}_{it}'$ refers to the coefficients associated with $\boldsymbol{z}_t'$ in the $i$th equation and $\tilde{v}_{ij, t}$ denotes the  dynamic regression coefficients  on the $j$th
contemporaneous value in the $i$th equation.  Note that for the first equation we have $\tilde{\boldsymbol{z}}_{1t}= \boldsymbol{z}_t$ and  $\tilde{\boldsymbol{\beta}}_{1t}=\boldsymbol{\beta}_{1t}$. 
The law of motion of the $j$th element of $\tilde{\boldsymbol{\beta}}_{it}$ reads
\begin{equation}
\tilde{\beta}_{ij,t} = \tilde{\beta}_{ij,t-1} + e_{ij,t} \quad e_{ij,t} \sim \mathcal{N}(0,\theta_{ij,t}).
\end{equation}
Hereby, $\theta_{ij,t}$ is defined analogously to \autoref{eq:threshold_1b}.

While not being order-invariant, this specific way of  stating the model  yields two significant computational gains.  First,  the matrix operations involved in estimating the latent state vector become computationally less cumbersome. Second, we can exploit parallel computing and estimate each equation simultaneously on a grid.

\subsection{Prior specification}
\label{sec:priors}

We impose a Normal-Gamma prior \citep{griffin2010inference} on each element of $\tilde{\boldsymbol{\beta}}_{i0}$, the initial state of the $i$th equation,
\begin{equation}
\tilde{\beta}_{ij,0}|\tau_{ij} \sim \mathcal{N}\left(0, \frac{2}{\lambda_i^2}  \tau^2_{ij}\right),~\tau^2_{ij} \sim \mathcal{G}(a_{i},a_{i}),
\end{equation}
for $i=1,\dots,m$ and $j=1,\dots, K_i$. Hereby, $\lambda_i^2$ and $a_{i}$ are hyperparameters and $\tau^2_{ij}$ denotes an idiosyncratic scaling parameter that applies an individual degree of shrinkage on each element of $\tilde{\boldsymbol{\beta}}_{i0}$. The hyperparameter $\lambda_i^2$ serves as an equation-specific shrinkage parameter that shrinks all elements of $\tilde{\boldsymbol{\beta}}_{i0}$ that belong to the $i$th equation towards zero while the local shrinkage parameters $\tau_{ij}$ provide enough flexibility to also allow for non-zero values of $\tilde{\beta}_{ij, 0}$ in the presence of a tight equation-specific prior. 

For the equation-specific scaling parameter $\lambda_i^2$ we impose a Gamma prior,
$
\lambda_i^2 \sim \mathcal{G}(b_0,b_1),
$
with $b_0$ and $b_1$ being hyperparameters chosen by the researcher. In typical applications we specify $b_0$ and $b_1$ to render this prior effectively non-influential. 

If the  innovation variances  of the observation equation are assumed to be constant over time, we impose a Gamma prior on  $\sigma_i^{-2}$ with hyperparameters $c_0$ and $c_1$, i.e.,~$\sigma_i^{-2} \sim \mathcal{G}(c_0, c_1)$.  By contrast,  if  stochastic volatility is introduced we follow \cite{kastner2014ancillarity} and impose a normally distributed prior on $\mu_i$ with mean zero and variance $100$, a Beta prior on $\rho_i$  with $(\rho_i+1)/2\sim \mathcal{B}(a_\rho,b_\rho)$, and a Gamma distributed prior on $\zeta_i \sim \mathcal{G}(1/2, 1/(2B_\zeta))$.

In the paper at hand, we only estimate the slab variance $\vartheta_{ij, 1}$ from the data and set $\vartheta_{ij,0} = \xi \times \hat{\vartheta}_{ij}$,
where $\hat{\vartheta}_{ij}$ denotes the variance of the OLS estimate for automatic scaling which we treat as a constant specified a priori.
The multiplier $\xi$ is set to a fixed constant close to zero, effectively turning off any time-variation in the parameters. As long as $\vartheta_{ij,0}$ is  not chosen too large, the specific value of the spike variance proves to be rather non-influential in the empirical applications that follow. 
Note that in principle, also the spike variance $\vartheta_{ij,0}$
could be estimated from the data and a suitable shrinkage prior could be employed to push $\vartheta_{ij,0}$ towards zero.

We use an Inverse-Gamma prior on the slab innovation variances in the state specification, i.e.,~$
\vartheta_{ij,1}^{-1} \sim \mathcal{G}(r_{ij,0}, r_{ij, 1})$ for $i=1,\dots,m$ and  $j=1,\dots,K_i$.\footnote{Of course, it would also be possible to use a (restricted) Gamma prior on $\vartheta_{ij,1}$ in the spirit of \cite{fruhwirth2010stochastic}. However, we have encountered some issues with such a prior if the number of observations in the regime associated with $\hat{s}_{ij,t}=1$ is small. This stems from the fact that the corresponding conditional posterior distribution is generalized inverse Gaussian, a distribution that can be heavy tailed and under certain conditions leads to excessively large draws of $\vartheta_{ij,1}$.} 
Again, $r_{ij, 0}$ and $r_{ij, 1}$ denote scalar hyperparameters. This choice implies that we artificially bound $\vartheta_{ij,1}$ away from zero, implying that in the upper regime we do not exert strong shrinkage. This is in contrast to a standard time-varying parameter model, where this prior is usually set rather tight to control the degree of time variation in the parameters \citep[see, e.g.,][]{primiceri2005time}. Note that in our model the degree of time variation is governed by the thresholding mechanism instead.

Finally, the prior specification of the baseline model is completed by imposing a uniform distributed prior on the thresholds,
\begin{equation}
d_{ij} \sim \mathcal{U}(\pi_{ij,0}, \pi_{ij,1}), \quad j=1,\dots,K_i. \label{eq:priorthresholds}
\end{equation}
Here, $\pi_{ij,0}$ and $\pi_{ij,1}$ denote the boundaries of the prior that have to be specified carefully. In our examples, we use $\pi_{ij, 0} = 0.1 \times \sqrt{\vartheta_{ij,1}}$ and $\pi_{ij,1} =  1.5 \times \sqrt{\vartheta_{ij,1}}$. This prior bounds the thresholds away from zero, implying that a certain amount of shrinkage is always imposed on the autoregressive coefficients. Setting $\pi_{ij,0}=0$ for all $i,j$ would also be a feasible option but we found in simulations that being slightly informative on the presence of a threshold improves the empirical performance of the proposed model markedly. It is worth noting that even under the assumption that $\pi_{0j}>0$, our framework performs well in simulations where the data is obtained from a non-thresholded version of our model. This stems from the fact that in a situation where parameters are expected to evolve smoothly over time, the average period-on-period change of $\beta_{ij,t}$ is small, implying that $0.1 \times \sqrt{\vartheta_{ij,1}}$ is close to zero and the model effectively shrinks small parameter movements to zero. 


\subsection{Posterior simulation}
\label{sec:mcmc}
We sample from the joint posterior distribution of the model parameters by utilizing an MCMC algorithm. Conditional on the thresholds $d_{ij}$, the remaining parameters can be simulated in a straightforward fashion. After initializing the parameters using suitable starting values we iterate between the following six steps.

\begin{enumerate}
\item We start with equation-by-equation simulation of the full history $\{\tilde{\boldsymbol{\beta}}_{it}\}_{t=0,1,\dots,T}$ for each $i$
by means of a standard forward filtering backward sampling algorithm \citep{carter1994gibbs, fruhwirth1994data} while conditioning on the remaining parameters of the model.

\item The reciprocals of the slab innovation variances, $\vartheta^{-1}_{ij, 1}$, $i=1,\dots,m$, $j=1,\dots,K_i$, have conditional density
\[
p(\vartheta^{-1}_{ij, 1}|\bullet)=p(\vartheta^{-1}_{ij, 1}|d_{ij},\tilde{\boldsymbol{\beta}}_{ij,0:T}) \propto p(\tilde{\boldsymbol{\beta}}_{ij,0:T}|\vartheta^{-1}_{ij, 1},d_{ij})p(d_{ij}|\vartheta^{-1}_{ij, 1})p(\vartheta^{-1}_{ij, 1}),
\]
where $\tilde{\boldsymbol{\beta}}_{ij,0:T} = (\tilde\beta_{ij,0},\dots,\tilde\beta_{ij,T})'$. This is a Gamma distribution, i.e.,
\begin{equation}
\vartheta^{-1}_{ij,1}|\bullet \sim \mathcal{G}\left(r_{ij,0} + \frac{T_{ij,1}}{2} + \frac{1}{2},r_{ij,1}+\frac{\sum_{t=1}^{T}\hat{s}_{ij,t}(\tilde{\beta}_{ij,t}-\tilde{\beta}_{ij,t-1})^2}{2}\right),
\end{equation}
with $T_{ij,1}=\sum_{t=1}^T \hat{s}_{ij, t}$ denoting the number of time periods that feature time variation in the $j$th parameter and the $i$th equation.

\item Combining the Gamma prior on $\tau_{ij}^2$ with the Gaussian likelihood yields a Generalized Inverted Gaussian (GIG) distribution
\begin{equation}
\tau_{ij}^2|\bullet \sim \mathcal{GIG}\left(a_{i}-\frac{1}{2}, \tilde{\beta}_{ij, 0}^2, a_{i} \lambda_i^2\right),
\end{equation}
where the density of $\mathcal{GIG}(\kappa, \chi,\psi)$ is proportional to 
$
z^{\kappa-1} \exp\left\lbrace - \left( \chi/z+\psi z\right)\right/2\rbrace.
$
To sample from this distribution, we use the R package GIGrvg \citep{GIGrvg} implementing the efficient rejection sampler proposed by \cite{hoermann2013generating} for each $i$ and $j$.

\item For each $i$, the global shrinkage parameter $\lambda_i^2$ is sampled from a Gamma distribution given by
\begin{equation}
\lambda_i^2| \bullet \sim \mathcal{G}\left(b_0+a_i K_i, b_1+\frac{a_i}{2}\sum_{j=1}^{K_i} \tau^2_{ij}\right).
\end{equation}

\item We update the thresholds by applying $K_i$ Griddy Gibbs steps \citep{ritter1992facilitating} per equation. Due to the structure of the model, the conditional distribution of $\Delta\tilde{\boldsymbol{\beta}}_{ij,1:T}$ is multivariate Gaussian, i.e.
\begin{equation}
 p\left(\Delta\tilde{\boldsymbol{\beta}}_{ij,1:T} | d_{ij}, \vartheta_{ij,0}, \vartheta_{ij,1}\right) \propto \prod_{t=1}^T \frac{1}{\sqrt{2 \pi \theta_{ij, t} }} \exp \left\lbrace -\frac{(\tilde{\beta}_{ijt}-\tilde{\beta}_{ij,t-1})^2}{2 \theta_{ij, t}}\right\rbrace.
\end{equation}
This expression can be straightforwardly combined with the prior in \autoref{eq:priorthresholds} to evaluate the conditional posterior of $d_{ij}$ at a given candidate point. The procedure is repeated over a fine grid of values that is determined by the prior and an approximation to the inverse cumulative distribution function  of the posterior is constructed.\footnote{In all applications, we use an evenly spaced grid that contains $150$ grid points.} Finally, this approximation is used to perform inverse transform sampling. 
\item The coefficients of each of the the log-volatility equations and the corresponding histories of the log-volatilities are sampled as in \cite{kastner2014ancillarity} through the R package \texttt{stochvol} \citep{kastner2016dealing}. Under homoscedasticity, $\sigma_i^{-2}$ is simulated from $\sigma_i^{-2}|\bullet \sim \mathcal{G}\left(c_0+T/2, c_1+{\sum_{t=1}^T (y_{it}-\boldsymbol{z}_{it}' \tilde{\boldsymbol{\beta}}_{it})^2}/{2}\right).$
\end{enumerate}
After obtaining an appropriate number of draws, we discard the burn-in and base our inference on the remaining draws from the joint posterior. 

In comparison with standard TVP-VARs, Step (5) is the only additional MCMC step needed to estimate the proposed TTVP model. Moreover, note that this update is computationally cheap, increasing the amount of time needed to carry out the analysis conducted in Section~\ref{sec:application2} by around five percent. For larger models (i.e.,\ with $m$ being around $15$) this step becomes slightly more intensive but, relative to the additional computational burden introduced by applying the FFBS algorithm in Step (1), its costs are still comparably small relative to the overall computation time needed. 

In the applications that follow, we draw $30\,000$ samples and discard the first $25\,000$ draws as burn-in. We found that mixing and convergence properties of our proposed algorithm are similar to standard Bayesian TVP-VAR estimators. In other words, the sampling of the thresholds does not seem to substantially increase the autocorrelation of the remaining MCMC draws. Concerning the threshold parameters themselves, we also observe quick mixing. Appendix~\ref{sec: convergence} provides some selected convergence criteria for the application to US macroeconomic data.

\section{An illustrative example}
\label{sec:illustration}

In this section we illustrate our approach by means of a rather stylized example that emphasizes how well the mixture innovation component for the state innovations performs when applied to different simulated scenarios.

For demonstration purposes it proves to be convenient to work with the following simple data generating process (DGP) with $K=1$ and $m=1$:
\begin{align*}
 y_t &= x_{11,t}' \beta_{11,t} + u_{1t}, ~u_{1t} \sim \mathcal{N}(0, 0.1^2), \\
\beta_{11,t} &= \beta_{11,t-1} + e_{11,t}, ~e_{11,t} \sim \mathcal{N}(0,s_{11,t} \times 0.1^2),
\end{align*}
where $s_{11,t} \in \{0,1\}$ is chosen to yield paths which are characterized by no ($s_{11,t} \equiv 0$ for all $t$), few, and many breaks, as well as a standard TVP DPG ($s_{11,t} \equiv 1$ for all $t$). Finally, independently for all $t = 1,\dots,500$, we generate $x_{11,t} \sim \mathcal{U}(-1,1)$.

In order to assess how different models perform in recovering the latent processes, we run a standard TVP model, a mixture innovation model estimated using the algorithm outlined in \cite{gerlach2000efficient}, and our TTVP model. To ease comparison between the models we impose a similar prior setup for all models. Specifically, for $\sigma_1^{-2}$ we set $c_0=0.01$ and $c_1=0.01$, implying a rather vague prior. For the shrinkage part on $\beta_{11,0}$ we set $\lambda_{1}^2 \sim \mathcal{G}(0.01,0.01)$ and $a_1 = 0.1$, effectively applying heavy shrinkage on the initial state of the system. The prior on $\vartheta_{11,1}$ is specified as in \cite{nakajima2013bayesian}, i.e., $\vartheta^{-1}_{11,1} \sim \mathcal{G}(3,0.03)$. To complete the prior setup for the TTVP model we set $\pi_{11,0}=0.1\times \sqrt{\vartheta_{11,1}}$ and $\pi_{11,1}=1.5\times \sqrt{\vartheta_{11,1}}$. Finally, $\xi$ is set equal to $0.01$.

\begin{figure}[t]
\includegraphics[width=.5\textwidth, trim=0 0 0 22, clip=true]{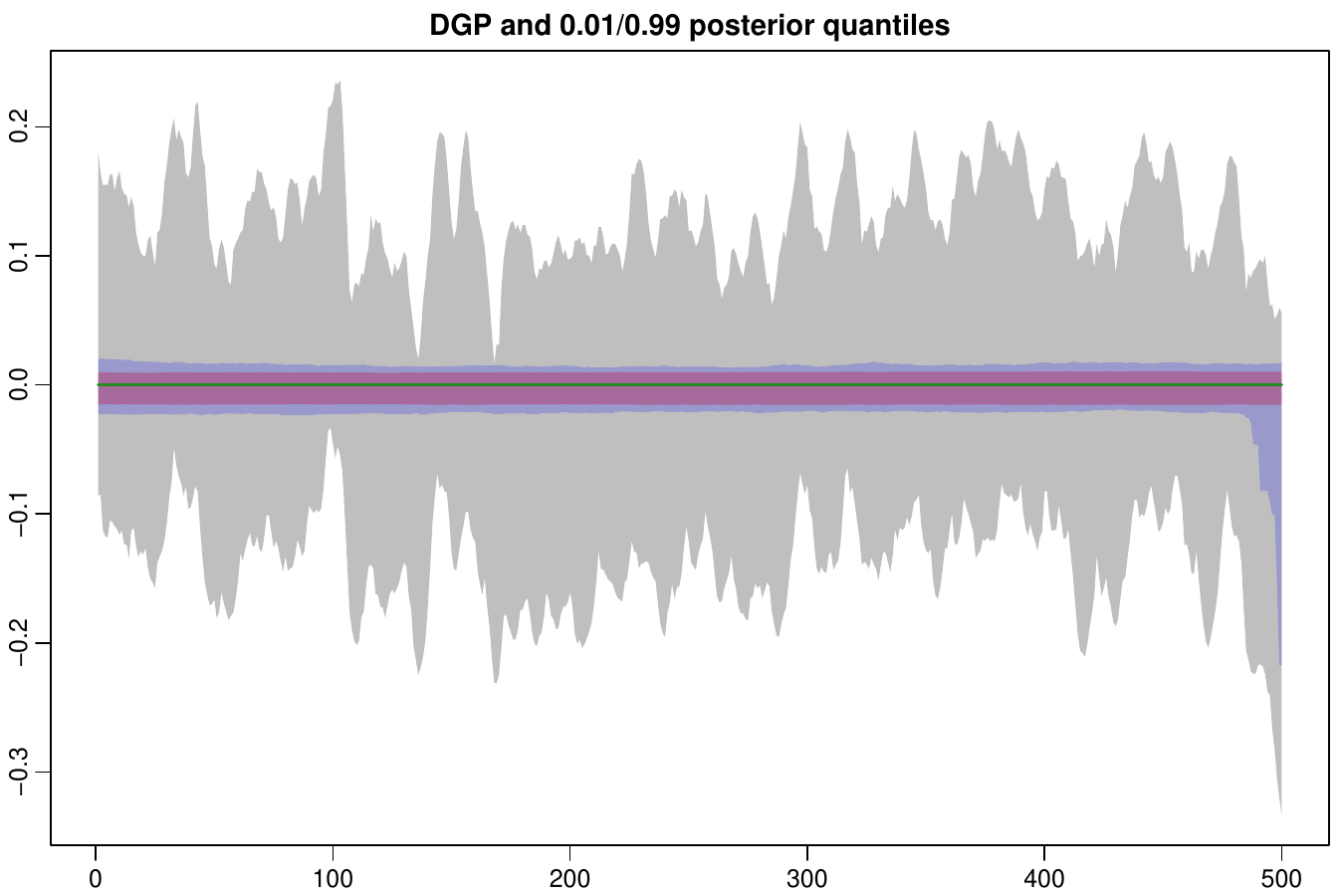}
\includegraphics[width=.5\textwidth, trim=0 0 0 22, clip=true]{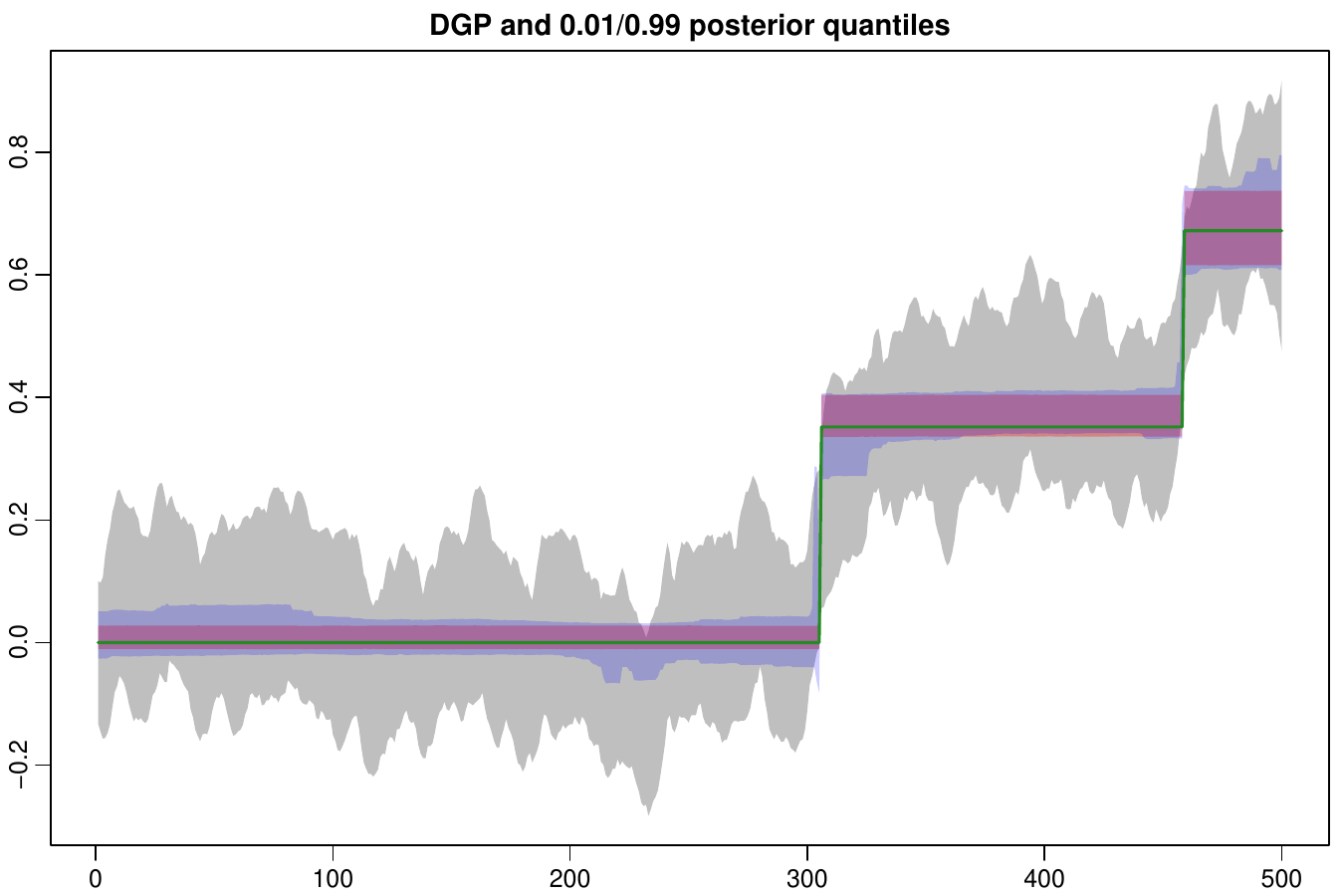}\\
\includegraphics[width=.5\textwidth, trim=0 0 0 22, clip=true]{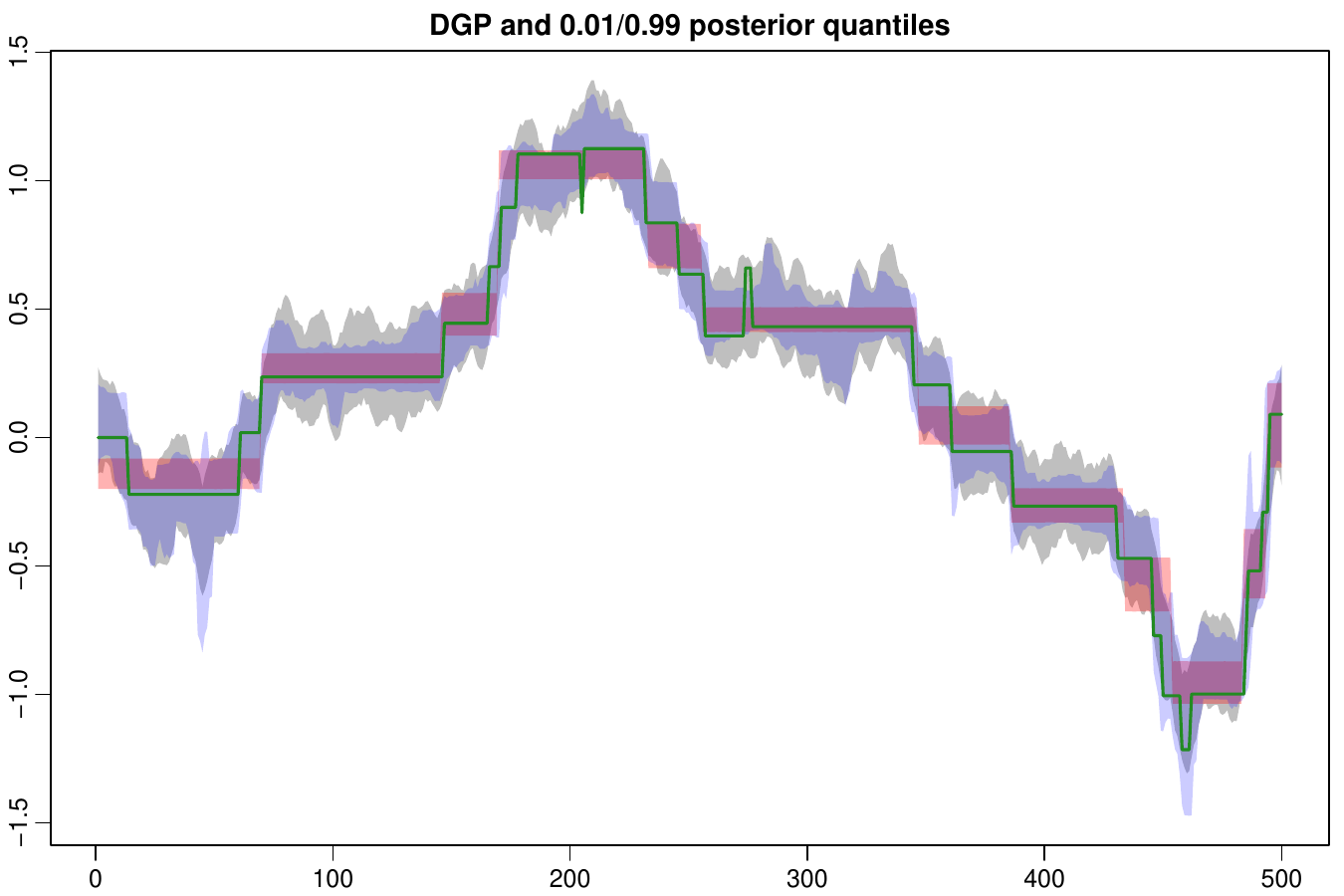}
\includegraphics[width=.5\textwidth, trim=0 0 0 22, clip=true]{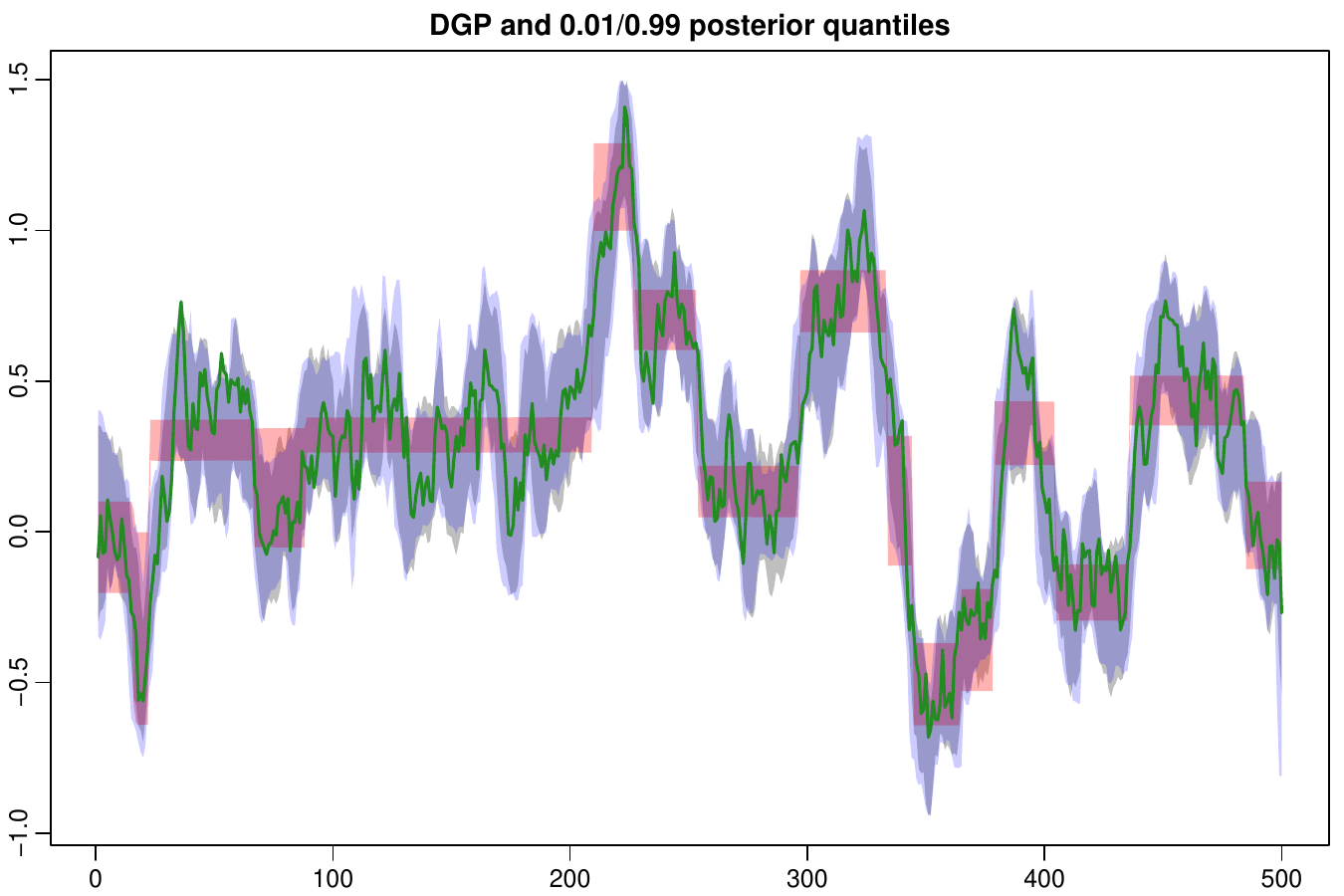}
\caption{Evolution of the actual state vector (sold green) along with the 98 percent posterior credible intervals of the TVP model (gray shaded area), the TTVP model (red shaded area) and a standard mixture innovation model (blue shaded area).}
\label{fig:examples}
\end{figure}

Figure \ref{fig:examples} displays the evolution of the 98 percent posterior credible intervals of the latent state vectors for standard TVP models (gray), mixture innovation models (blue) and TTVP models (red) along with the actual evolution of the state vector (green). Each panel of \autoref{fig:examples} is based on a single realization from the data generating process.

At least three interesting findings emerge. First, note that our approach captures parameter movements rather well,  signaling large jumps for virtually  all time points that feature a structural break in the corresponding parameter. The TVP model also tracks the actual movement of the states well but does so with much more high frequency variation. This is a direct consequence of the inverted Gamma prior on the state innovation variances that bound $\vartheta_{11,1}$ artificially away from zero, irrespective of the information contained in the likelihood \citep[see][for a general discussion of this issue]{fruhwirth2010stochastic}.  

Second, investigating posterior uncertainty reveals that our approach succeeds in shrinking the posterior variance. This is due to the fact that in periods where the true value of $\beta_{11,t}$ is constant, our model successfully assumes that the estimate of the coefficient at time $t$ is also constant, whereas the TVP model imposes a certain amount of time variation. This generates additional uncertainty  that inflates the posterior variance, possibly leading to imprecise inference. The standard mixture innovation model is also capable of reducing uncertainty effectively, but at a much larger computational cost as compared to our proposed modeling approach.

Third, contrasting the findings of the TTVP specification with the results obtained from a standard mixture innovation model provides some evidence that our approximation works rather well if the DGP is characterized by not more than a moderate amount of jumps. By contrast, if the DGP features sustained movement, our approach pushes the majority of the high frequency variation to zero. This stems from the fact that we are slightly informative on the specific value of the threshold, effectively ruling out the case where the threshold is zero. Notice, however, that while the posterior distribution of the standard mixture innovation model converges to the posterior distribution of the TVP specification, the corresponding posterior uncertainty also rises sharply. We conjecture that especially in forecasting applications, capturing large swings in the parameters  could be sufficient to adequately describe key relations in the data while the reduction in parameter uncertainty could ultimately  lead to more precise predictions.

To sum up, the TTVP model detects change points in the parameters in situations where the actual number of breaks is small, moderate and large. In situations where the DGP suggests that the actual threshold equals zero, our approach still captures most of medium to low frequency noise but shrinks small movements that might, in any case, be less relevant for econometric inference.

\section{Forecasting the US term structure of interest rates}\label{sec:application1}
The first empirical application deals with predicting the US term structure of interest rates. Forecasting the term structure appears to be an important task for policy makers and practitioners  alike. Central banks are interested in how their policy interventions impact the different segments of the term structure and how these movements transmit into the real economy. From a forecasting perspective, precise predictions are necessary for various tasks such as active portfolio management, risk  management as well as general policy analysis. 

Several attempts have been made to predict the term structure of interest rates with a wide range of different models \citep[see, among many others,][]{diebold2006forecasting, monch2008forecasting, monch2012term, favero2012term, carriero2012forecasting, carriero2014no, byrne2017forecasting}. The majority of these contributions, however, focuses exclusively on evaluating point predictions while ignoring higher moments of the underlying predictive distribution. In addition, most studies typically assume that the parameters of the model are constant over time. Some recent exceptions are \cite{bianchi2009great, mumtaz2009time, koopman2010analyzing, carriero2014no, byrne2017forecasting}.  In the paper at hand, we apply the TTVP model to predict the term structure of interest rates and benchmark it to various competing model specifications. 

\subsection{Data overview, model specification, and design of the forecasting exercise}
We use  monthly Fama-Bliss zero coupon yields  obtained from the Chicago Booth Center for Research in Security prices (CRSP) database as well as  the dataset described in \cite{gurkaynak2007us}. The data spans the period from 1960:M01 to 2014:M12 and  the maturities included are 1, 2, 3, 4, 5, 7, and 10 years.\footnote{The data for the maturities one up to five years are based on the CRSP data while the data for maturities seven and ten years are taken from the \cite{gurkaynak2007us} database.}  Moreover, we include $p=3$ lags of the endogenous variables.
The prior setup is similar to the one adopted in the previous section. 
More specifically, for all applicable $i$ and $j$, we use the following values for the hyperparameters. For the shrinkage part on the initial state of the system, we again set $\lambda_i^2 \sim \mathcal{G}(0.01,0.01)$ and $a_i = 0.1$.  For the parameters of the log-volatility equation we use $\mu_i \sim \mathcal{N}(0, 10^2)$, $\frac{\rho_i+1}{2} \sim \mathcal{B}(25,5)$, and $\zeta_i \sim \mathcal{G}(1/2, 1/2)$. The prior on the thresholds are set equal to $\pi_{ij, 0}=0.1 \times \sqrt{\vartheta_{ij,1}}$ and $\pi_{ij, 1}=1.5 \times \sqrt{\vartheta_{ij,1}}$. Finally, we assess the impact of different choices for the prior on $\vartheta_{ij, 1}$ as well as $\xi$ in \autoref{tab: forecasting_results}.

Our forecasting design is recursive. For an initial estimation period, in our case 1960:M01 to $T_0=\text{1999:M08}$, we compute predictions for the next three months
via Monte Carlo integration. More concretely, for each of the $l = 1,\dots,5000$ MCMC draws from the posterior distribution, we start by predicting $\bm \beta_{T_0+n}^{(l)}$ for $n=1,2,3$. This is achieved by first drawing an indicator $s^{(l)}_{ij, T_0+n}$ for all $i, j$ from a Bernoulli distribution with success probability ${T_{ij,1}^{(l)}}/{T_0}$. Using this indicator, we compute  $\theta^{(l)}_{ij, T_0+n}$ and thus infer whether the predicted change is effectively zero or not. This information enables us to recursively calculate $\bm{\beta}_{T_0+n}^{(l)}$ utilizing the state evolution in \autoref{eq:states1}. Next, we construct $\bm V_{T_0+n}^{(l)}$ through the corresponding elements in $\bm \beta_{T_0+n}^{(l)}$ and then predict the log-volatilities using \autoref{eq: logvolas} to compute $\bm{\Sigma}_{T_0+n}^{(l)}$. Finally, we obtain predictions for $\bm y_{T_0+n}$ by drawing from $\mathcal{N}\!\left(\bm{x}_{T_0+n}'\bm\beta_{T_0+n}^{(l)}, \bm{\Sigma}_{T_0+n}^{(l)}\right)$.

After obtaining these, we expand the initial estimation sample by one month and repeat this procedure until the end of the sample is reached.  This yields a sequence of $184 \times 3$ predictive densities. Forecasts are then evaluated using  log predictive scores (LPSs)  for a model of interest and a benchmark model. The LPS is a widely used metric to measure density forecast accuracy \citep[see, e.g.,][]{geweke2010comparing}. 

\subsection{Competing models}
As the benchmark model, we use a TVP-VAR with SV employing the prior setup described in \cite{primiceri2005time}. We, moreover,  include three additional constant parameter VAR models, namely a  Minnesota-type VAR \citep{Doan1984},\footnote{This specific implementation follows \cite{koop2010bayesian} but, following \cite{giannone2012prior}, estimates the hyperparameters using two Metropolis Hastings steps.}  a Normal-Gamma (NG) VAR \citep{Huber2017} and a VAR coupled with a stochastic search variable selection (SSVS) prior \citep{george2008bayesian}. For the SSVS VAR, we set the scaling parameters associated with the two Gaussian mixture components of the prior using the semi-automatic approach described in \cite{george2008bayesian}. This implies that the prior standard deviation for the slab component is ten times the corresponding OLS standard deviation while for the spike component it is one tenth of OLS standard deviation.

Moreover, and given its success in forecasting the term structure, we also benchmark our unrestricted multivariate model specifications with the model proposed in \cite{diebold2006forecasting} based on the three factor Nelson Siegel (NS) framework \citep{nelson1987parsimonious}. The NS approach imposes a factor structure on the yields,
\begin{equation}
i_t(\upsilon) = L_t + \frac{1-e^{-\upsilon \alpha}}{\upsilon \alpha} S_t + \left(\frac{1-e^{-\upsilon \alpha}}{\upsilon \alpha} - e^{-\upsilon \alpha}\right) C_t +m_t(\upsilon). \label{eq: NS_eq}
\end{equation}
Here, $i_t ( \upsilon)$ denotes the yield at maturity $\upsilon$, $L_t$ is a factor that controls the level, $S_t$ determines the slope, and $C_t$ represents the curvature factor of the yield curve. Moreover, we let $m_t(\upsilon)$ denote a pricing error. The parameter $\alpha$ controls the shape of the factor loadings in \autoref{eq: NS_eq} and is set to $\alpha=0.0609$ to maximize the loading on $C_t$ \citep[for a discussion of this particular choice, see][]{diebold2006forecasting}.  In what follows, we estimate the latent factors $L_t, S_t$, and $C_t$ by OLS.  These factors are then included in $\bm y_t = (L_t, S_t, C_t)'$ and a VAR with a Minnesota prior is estimated (labeled NS-VAR). Moreover, we also estimate a TTVP-NS-VAR model to assess whether allowing for time-variation in the state equation of the factors pays off.\footnote{For this specification, we use the prior setup described above and set $r_0=3, r_1=0.03$ and $\xi=0.001$.} Notice that since \autoref{eq: NS_eq} is a standard measurement equation, we forecast the yield curve by using the VAR state equation to compute the predictions in terms of the factors and then map it back to the yields using the factor loadings.

All models, including the random walk, feature stochastic volatility.  In order to assess the impact of different prior hyperparameters on $\vartheta_{ij, 1}$ and the impact of $\xi$, we estimate the TTVP model over a grid of meaningful values.

\subsection{Forecasting results}

\begin{table}[t!]
\caption{Log predictive Bayes factors relative to TVP-VAR over the hold-out period 1999:M09 to 2014:M12. Numbers greater than zero indicate that a given model outperforms the benchmark. 
The final column refers to the joint density forecasting performance, while the other columns refer to the univariate margins. Bold figures indicate the best performing model for each column and horizon.}\label{tab: forecasting_results}
\scalebox{0.75}{
\centering
\begin{tabular}{lrrrrrrrr}
  \toprule
  & 1Y & 2Y & 3Y & 4Y & 5Y & 7Y & 10Y & joint \\[.3em]
\multicolumn{1}{l}{}& \multicolumn{8}{c}{One-month-ahead}\\\midrule
TTVP-VAR: $\xi=0.1, r_0=3, r_1= 0.03$ 				& -107.1 & -92.1 & -67.0 & -41.3 & 11.5 & 22.0 & 10.8 & 2595.2 \\              
TTVP-VAR: $\xi=0.1, r_0=1.5, r_1= 1$ 					& -159.0 & -122.5 & -79.7 & -40.8 & 15.0 & 40.5 & 33.0 & 2713.0 \\             
TTVP-VAR: $\xi=0.1, r_0=0.001, r_1= 0.001$	   & -115.4 & -100.9 & -73.8 & -44.7 & 8.7 & 23.8 & 14.0 & 2696.2 \\                 
TTVP-VAR: $\xi=0.01, r_0=3, r_1= 0.03$ 					& -22.3 & -10.4 & 20.2 & 50.3 & 97.9 & 61.1 & 46.7 & 2927.3 \\             
TTVP-VAR: $\xi=0.01, r_0=1.5, r_1= 1$ 				& 25.3 & 21.0 & 40.2 & 64.9 & 106.1 & 70.7 & 55.3 & 2943.2 \\                  
TTVP-VAR: $\xi=0.01, r_0=0.001, r_1= 0.001$ 		 & -6.3 & 6.2 & 32.5 & 59.0 & 102.4 & 74.0 & 59.5 & 2899.7 \\                      
TTVP-VAR: $\xi=0.001, r_0=3, r_1= 0.03$ 					& 28.5 & 20.7 & 40.3 & 65.3 & 108.3 & 74.0 & 60.0 & \textbf{2976.1} \\              
TTVP-VAR: $\xi=0.001, r_0=1.5, r_1= 1$ 				& 28.2 & 21.5 & 40.4 & 65.4 & 107.9 & 71.7 & 57.8 & 2953.3 \\                  
TTVP-VAR: $\xi=0.001, r_0=0.001, r_1= 0.001$ 		 & 28.5 & 22.0 & \textbf{40.9} & \textbf{65.7} & 108.3 & 73.0 & 59.0 & 2952.9 \\
NS-TTVP-VAR 												& -26.6 & -17.5 & 20.5 & 57.7 & \textbf{110.1} & 101.0 & 82.8 & 2655.8 \\
 \midrule
NS-VAR 												& -61.6 & -39.9 & 8.8 & 53.1 & 109.8 & 125.0 & 119.3 & 2681.0 \\   
Minnesota-type VAR 		                               & 22.7 & 18.8 & 26.7 	& 3.6 & 4.5 & 138.0 & 128.7 & 2660.2 \\    
NG VAR 								& \textbf{35.4} & 22.5 & 19.2 & -45.0 & -50.6 & \textbf{141.9} & \textbf{136.1} & 2505.9 \\    
SSVS VAR													& -30.8 & -51.9 & -38.5 & -13.6 & 37.8 & 42.3 & 41.0 & 1308.8 \\       
Random walk 													& 35.3 & \textbf{22.8} & 31.6 & 14.8 & 20.1 & 141.0 & 135.0 & 2656.0 \\[1em]
    \multicolumn{1}{l}{}& \multicolumn{8}{c}{Three-months-ahead}\\\midrule 
TTVP-VAR: $\xi=0.1, r_0=3, r_1= 0.03$ 								& 	-279.4 & -225.4 & -181.0 & -164.7 & -163.8 & -175.1 & -148.8 & 625.2 \\     
TTVP-VAR: $\xi=0.1, r_0=1.5, r_1= 1$ 								&	 -330.0 & -257.7 & -194.8 & -162.2 & -160.4 & -164.2 & -123.5 & 809.4 \\        
TTVP-VAR: $\xi=0.1, r_0=0.001, r_1= 0.001$					& -283.3 & -229.3 & -182.2 & -161.2 & -160.6 & -168.0 & -137.9 & 786.8 \\               
TTVP-VAR: $\xi=0.01, r_0=3, r_1= 0.03$ 								& -82.7 & -34.3 & 19.2 & 40.2 & 35.6 & -14.6 & 16.3 & 1214.9 \\                 
TTVP-VAR: $\xi=0.01, r_0=1.5, r_1= 1$ 							& 8.3 & 36.4 & 73.8 & 85.8 & 82.7 & 40.6 & 71.8 & 1293.2 \\                         
TTVP-VAR: $\xi=0.01, r_0=0.001, r_1= 0.001$ 	 				& -52.7 & 0.7 & 50.1 & 62.6 & 56.8 & 14.8 & 45.2 & 1194.1 \\                            
TTVP-VAR: $\xi=0.001, r_0=3, r_1= 0.03$ 							 & 14.8 & 36.4 & 72.5 & 87.2 & 81.7 & 41.1 & 71.9 & \textbf{1333.4} \\                       
TTVP-VAR: $\xi=0.001, r_0=1.5, r_1= 1$ 							& 15.5 & 39.0 & \textbf{75.1} & \textbf{88.6} & \textbf{84.3} & 42.8 & 73.9 & 1312.2 \\                        
TTVP-VAR: $\xi=0.001, r_0=0.001, r_1= 0.001$ 	 				& 16.0 & 39.1 & 75.0 & 88.5 & 84.2 & 43.3 & 74.3 & 1313.3 \\ 
NS-TTVP-VAR															& -30.6 & 5.0 & 56.8 & 84.3 & 82.0 & 39.1 & 67.6 & 1038.5 \\  \midrule                                      
NS-VAR 														& -74.4 & -20.2 & 43.7 & 80.1 & 83.3 & \textbf{46.5} & \textbf{80.6} & 1016.3 \\              
Minnesota-type VAR 													& -1.2 & 36.8 & 69.1 & 61.4 & 32.0 & 29.2 & 66.5 & 1005.7 \\            
NG VAR 													& \textbf{17.3} & \textbf{40.1} & 58.4 & 10.4 & -29.7 & 16.0 & 59.8 & 863.0 \\            
SSVS VAR 														& -86.5 & -74.2 & -44.9 & -27.7 & -28.2 & -55.8 & -20.3 & -614.7 \\     
Random walk 														& 17.0 & 39.4 & 68.4 & 54.5 & 17.9 & 28.7 & 69.1 & 975.7 \\             
   \bottomrule
\end{tabular}
}
\end{table}

Table \ref{tab: forecasting_results} shows the results for one-month- and three-months-ahead forecasts.  We start by considering the joint forecasting performance, provided in the rightmost column of Table \ref{tab: forecasting_results}, for the one-month ahead forecast horizon. Here we see that all models improve upon the standard TVP-VAR with SV by large margins. This clearly suggests that a standard TVP-VAR model equipped with inverted Gamma priors on the state innovation variances seems to overfit the data which, in turn, translates into a weak out-of-sample predictive performance. Comparing the predictive performance of our TTVP model across different choices for $\xi$ reveals that this parameter appears to be highly influential. If the hyperparameter is set too large, too little shrinkage is introduced and the forecasting performance deteriorates. Considering the different choices of $\xi$  shows that smaller values are typically accompanied by larger improvements in LPSs. Notice that the choice of $r_0$ and $r_1$ tends to play only a minor role compared to the scaling parameter $\xi$.

Relative to the remaining benchmark models, we find that using our TTVP approach appears to improve against all constant parameter VARs as well as the models based on the NS approach. The NS-VAR ranks second while the VAR with the hierarchical Minnesota prior ranks third. Using the TTVP framework in combination with the NS factors also produces predictions that are competitive to the remaining constant parameter specifications.  Contrasting the differences between the random walk and the Minnesota prior shows that both yield similar predictions. This is because the hierarchical Minnesota prior exerts strong shrinkage towards a random walk process. 

Considering the three-months-ahead predictions yields similar insights. The TTVP models continue to perform well, outperforming both the TVP-VAR with SV, the constant parameter VARs as well as the NS models. It is noteworthy that for multi-step forecasting, the SSVS VAR shows the weakest performance across all models considered, leading to a forecast performance that is even inferior compared to the benchmark.

Zooming into the results for the different maturities shows that especially for the short as well as for the long end of the yield curve, most constant parameter models seem to outperform their time-varying parameter competitors and the models based on the NS factors for the one-month ahead forecast horizon. For 1Y and 2Y maturities, the random walk as well as the NG VAR generate the most precise predictions, improving slightly upon the single best performing TTVP specification. The particularly strong predictive performance of the random walk for the short end of the yield curve has been found in several contributions on predicting interest rates \citep{diebold2006forecasting, carriero2012forecasting}. This finding, however, does not carry over to maturities between three and five years. There we find that our proposed framework as well as the NS models excel, clearly outperforming the competitors. Interestingly, for five year maturities we find that estimating a small-scale factor model with time-varying parameters and mixture innovations yields the strongest performance. Again, when considering multi-step-ahead forecasts we find a rather similar picture, with models that perform well in terms of one-step-ahead forecasting also doing well when three-step-ahead forecasts are considered.

\begin{figure}[t]
\centering
\includegraphics[width=\textwidth]{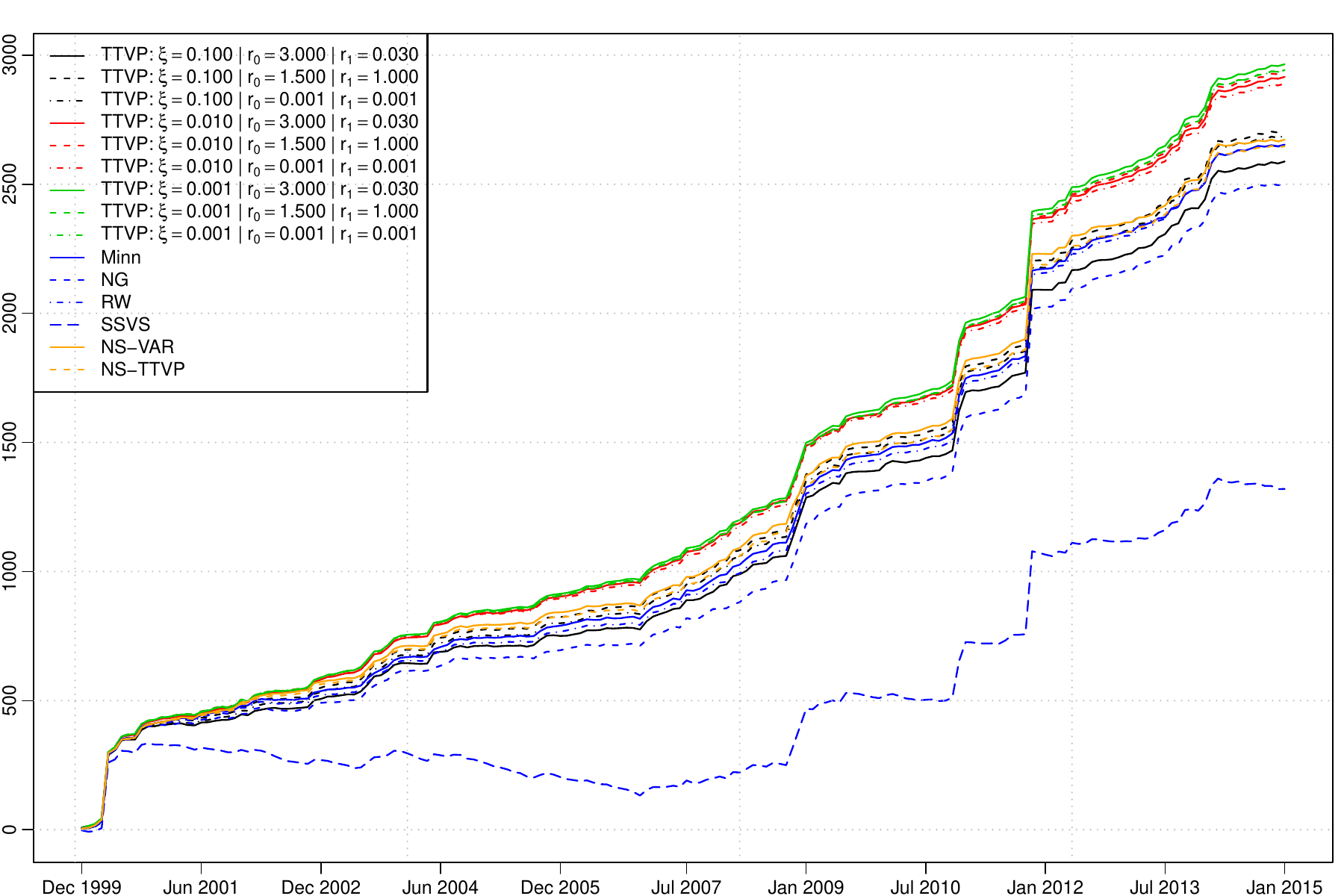}
\caption{One-month-ahead log predictive Bayes factor relative to a TVP-VAR model with SV.}
\label{fig:lps}
\end{figure}
Finally, we investigate whether the predictive performance varies with time. Figure \ref{fig:lps} displays the evolution of the one-step-ahead LPSs vis-\'{a}-vis the TVP-VAR with SV specification. At least two interesting patterns over time emerge. We find that during the financial crisis in 2008/2009, all competing models increase sharply against the benchmark specification. In addition, a pronounced jump in relative forecasting performance is also visible during the second half of 2011, a period characterized by the US debt ceiling crisis of 2011. Our conjecture is that this is driven by a) the ability to rapidly adjust regression coefficients and thus allow for changing transmission mechanisms in a flexible way and b) the fact that the TVP-VAR with SV overfits the data severely and  appears to be incapable of handling large shocks to the term  structure.

To sum up, this section highlights that using our TTVP approach generally pays off when used to predict the US term structure of interest rates. When considering the joint density forecasting performance, we find that this model framework improves sharply against the competing models used. If the forecaster's goal is to predict only certain segments of the yield curve, we find that for the short and the long end of the term structure, linear VARs with SV as well as the random walk with SV outperform our modeling approach. For three up to five year maturities, TTVP models with and without a factor structure on the yields outperform all remaining models.

\section{Structural breaks in US macroeconomic data}\label{sec:application2}
We complement the forecasting exercise by investigating the effects of a monetary policy shock. For that purpose we use a standard US macroeconomic data set, employed among others in \cite{smets2007shocks}, \cite{geweke2012prediction} and \cite{amisano2017prediction}. Data are on a quarterly basis, span the period from 1947Q2 to 2014Q4, and comprise the log differences of  consumption, investment, real GDP, hours worked,  consumer prices and real wages. Last, and as a policy variable, we include the Federal Funds Rate (FFR) in levels. The prior setup mirrors the one used in the preceding section but sets $\xi = 0.01/6$.\footnote{This value is based on running a forecasting exercise using this dataset and a hold-out period of 45 years. Specific results are available in the working paper version of this paper.} Following \cite{primiceri2005time}, we include $p=2$ lags of the endogenous variables.

In the next sections we start by proposing a global measure of time-variation in the VAR coefficients and then move on to analyze macroeconomic relations by means of impulse response analysis.
\subsection{Detecting time-variation in reduced form coefficients}
\begin{figure}[t!]
\begin{subfigure}{.327\textwidth}
\includegraphics[width=\textwidth, clip, trim = 30 40 30 40]{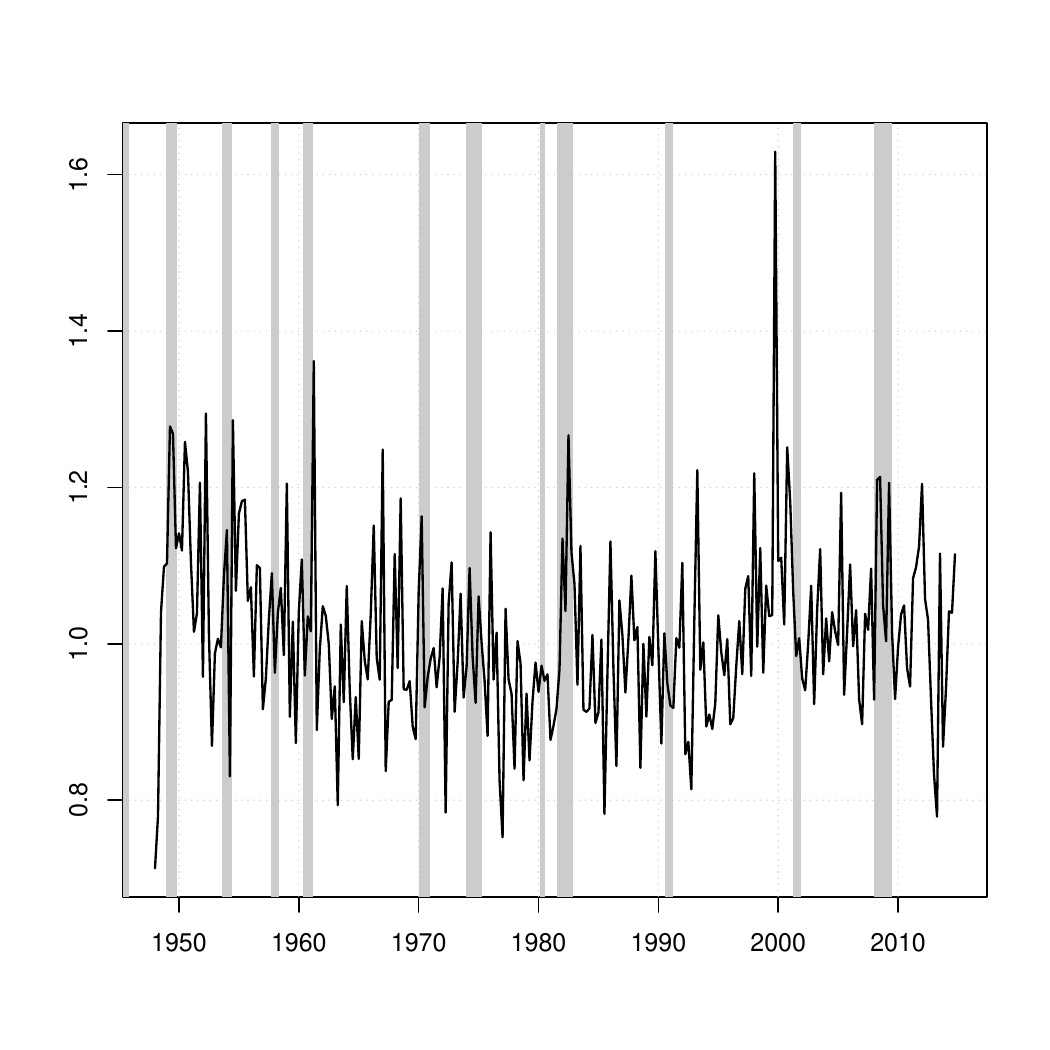}
\caption{Consumption}\label{fig:2a}
\end{subfigure}
\begin{subfigure}{.327\textwidth}
\includegraphics[width=\textwidth, clip, trim = 30 40 30 40]{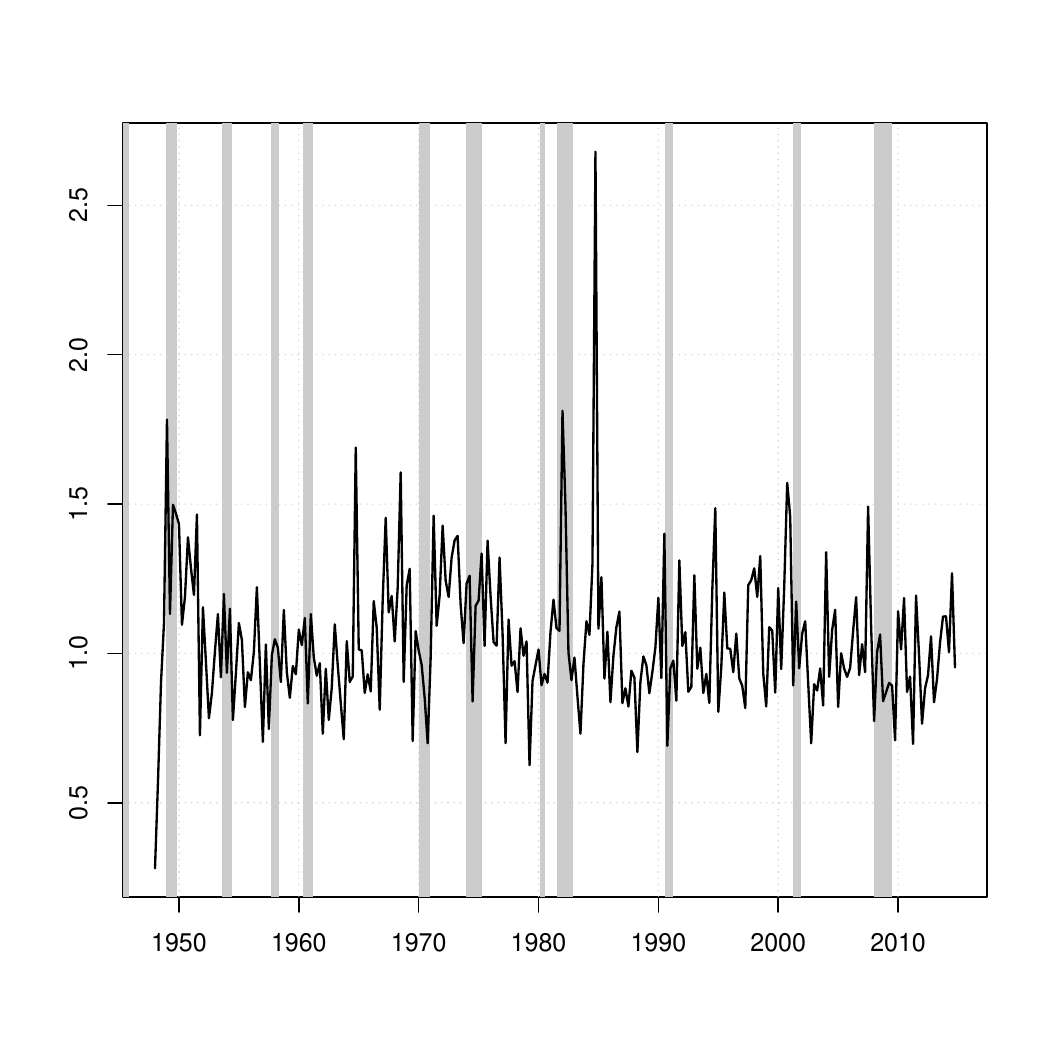}
\caption{Investment}\label{fig:2b}
\end{subfigure}
\begin{subfigure}{.327\textwidth}
\includegraphics[width=\textwidth, clip, trim = 30 40 30 40]{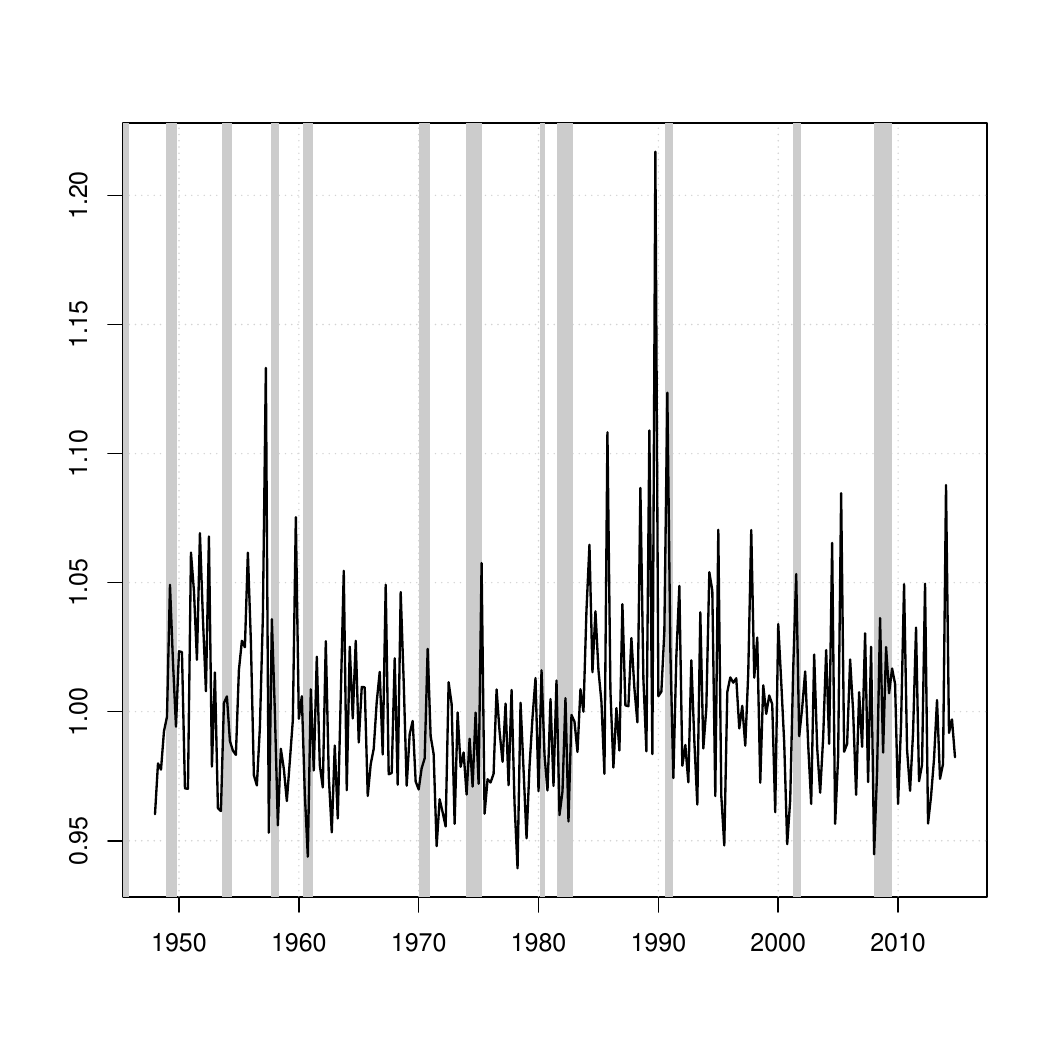}
\caption{Output}\label{fig:2c}
\end{subfigure}\\[.7em]
\begin{subfigure}{.327\textwidth}
\includegraphics[width=\textwidth, clip, trim = 30 40 30 40]{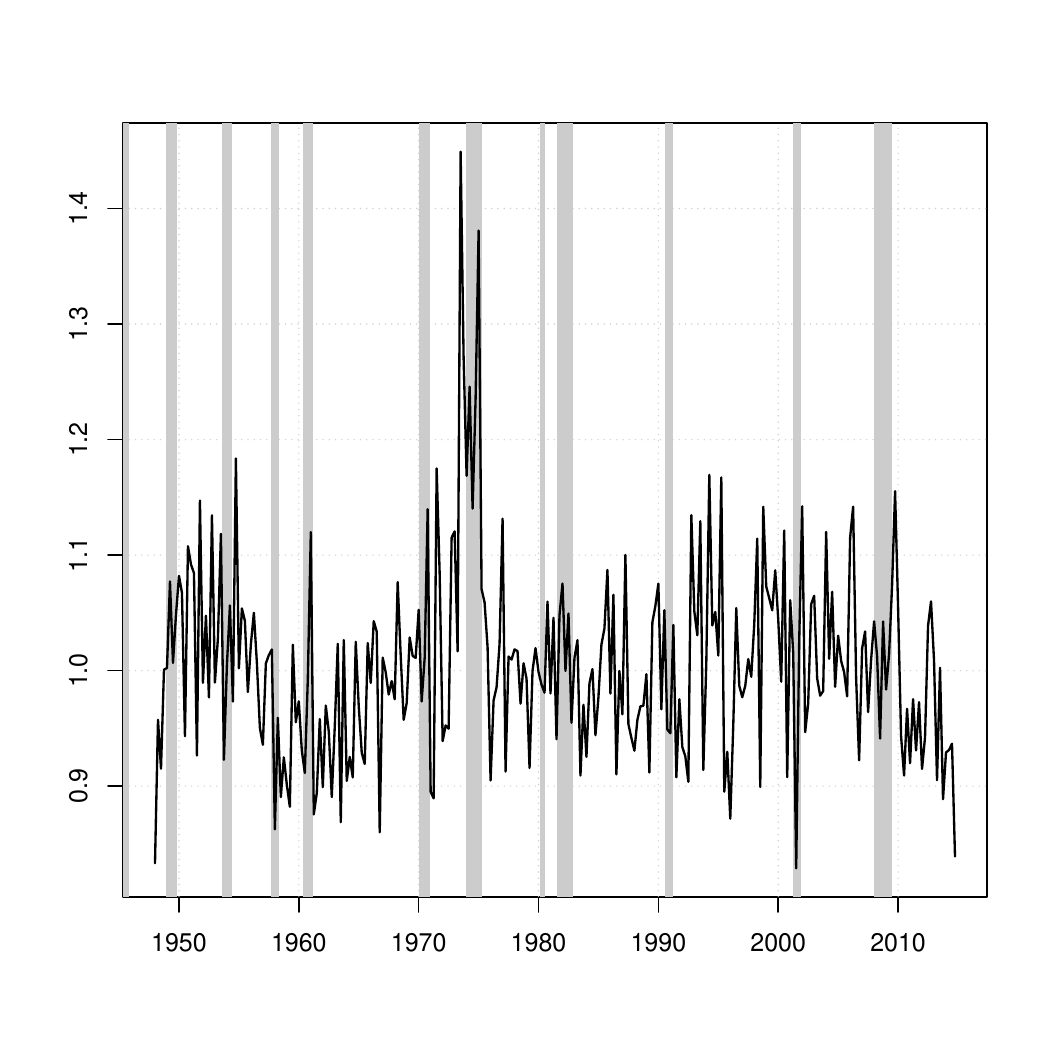}
\caption{Hours worked}\label{fig:2d}
\end{subfigure}
\begin{subfigure}{.327\textwidth}
\includegraphics[width=\textwidth, clip, trim = 30 40 30 40]{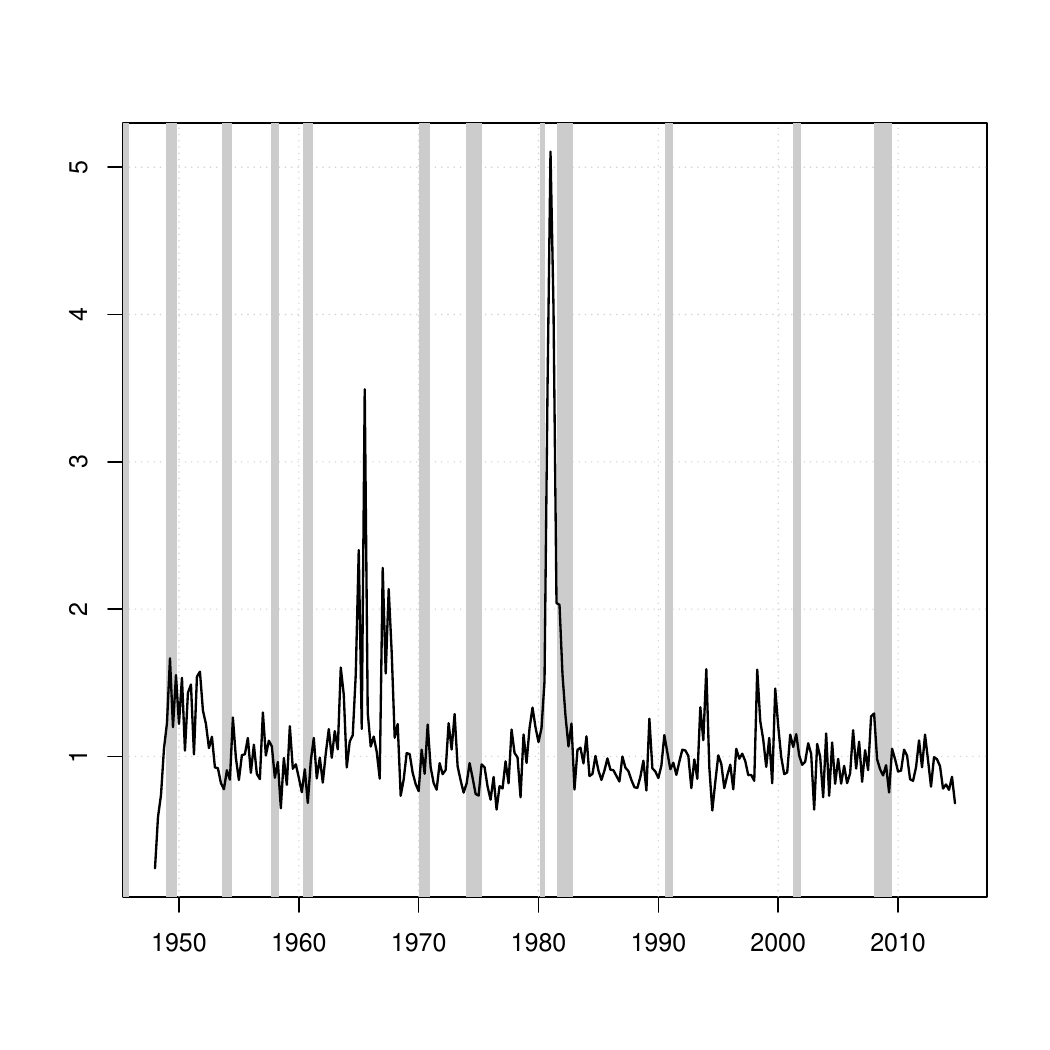}
\caption{Inflation}\label{fig:2e}
\end{subfigure}
\begin{subfigure}{.327\textwidth}
\includegraphics[width=\textwidth, clip, trim = 30 40 30 40]{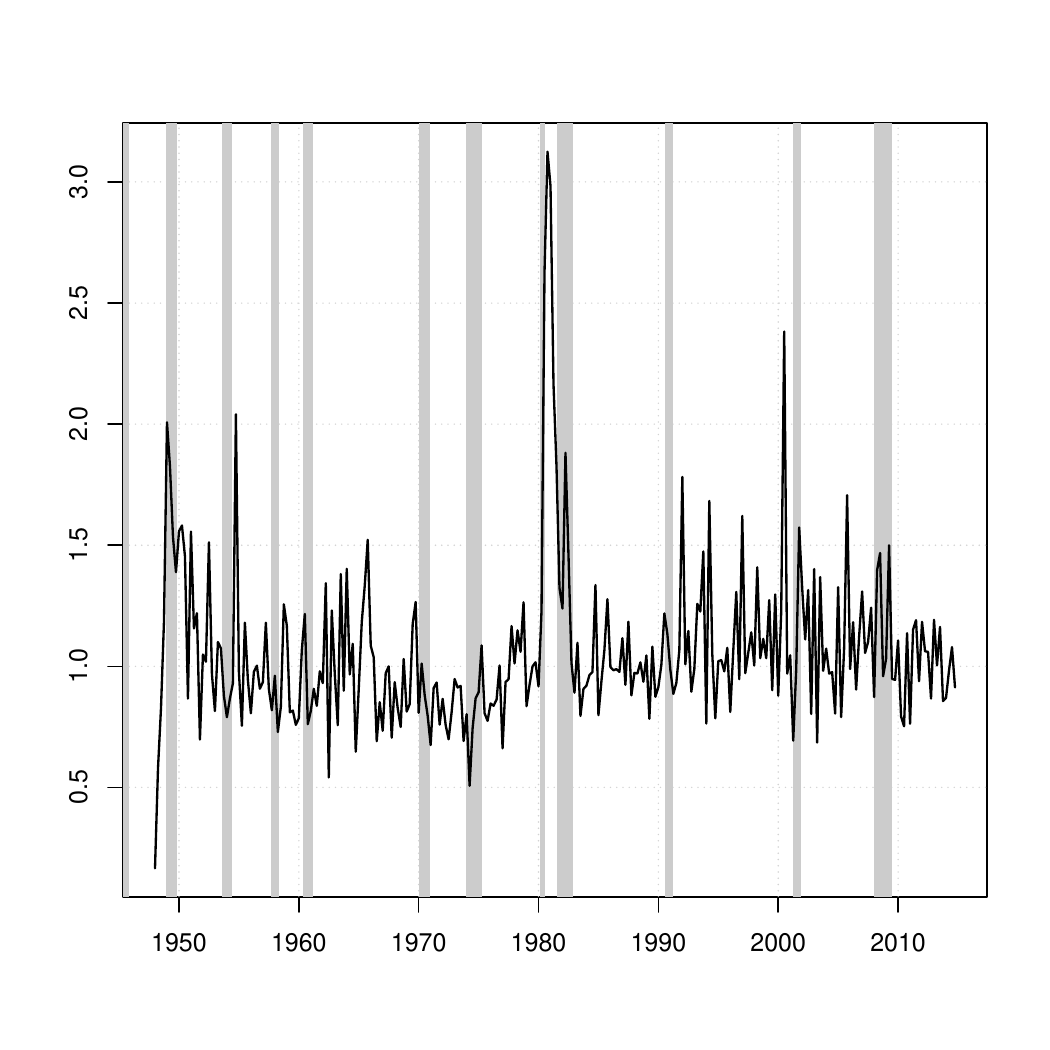}
\caption{Real wages}\label{fig:2f}
\end{subfigure}\\[.7em]
\centering
\begin{subfigure}{.327\textwidth}
\includegraphics[width=\textwidth, clip, trim = 30 40 30 40]{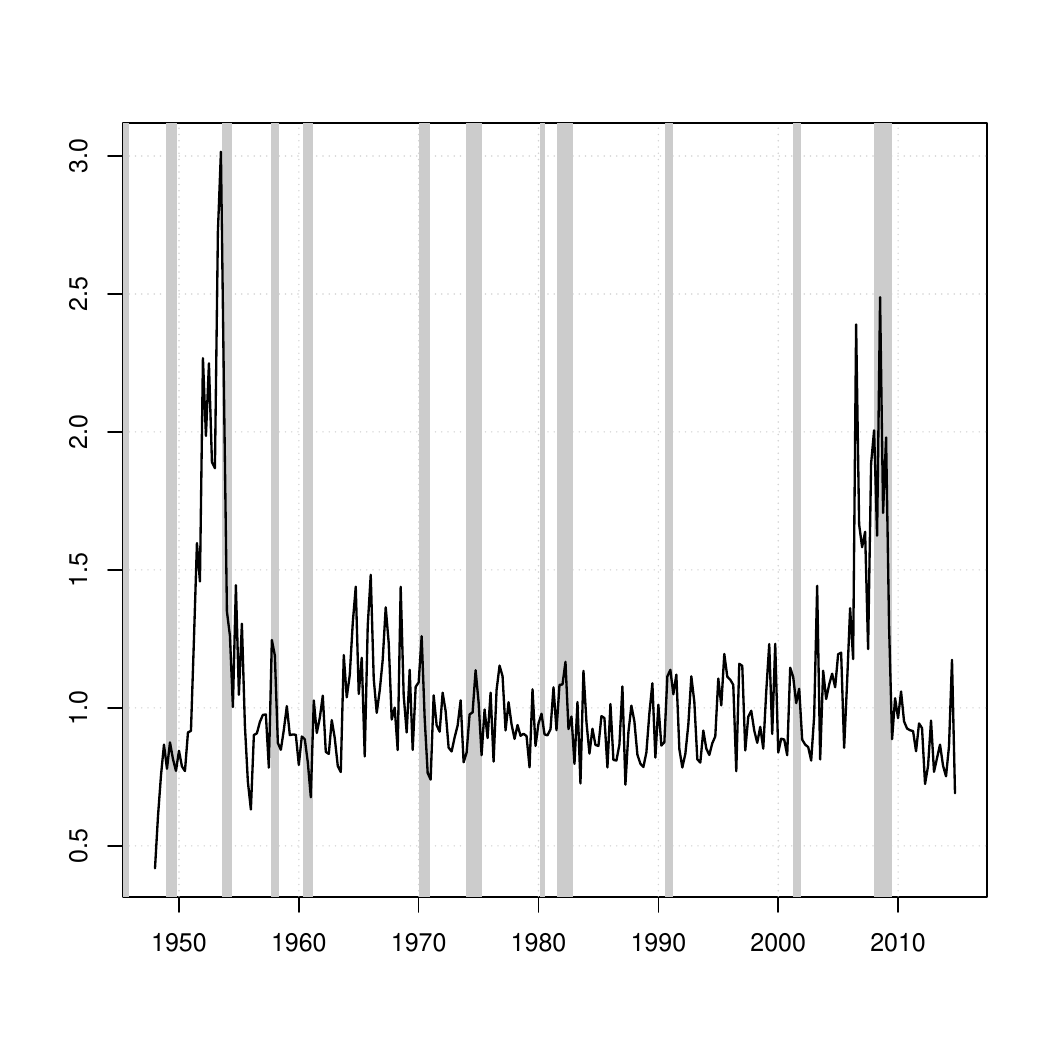}
\caption{FFR}\label{fig:2g}
\end{subfigure}
\begin{subfigure}{.327\textwidth}
\includegraphics[width=\textwidth, clip, trim = 30 40 30 40]{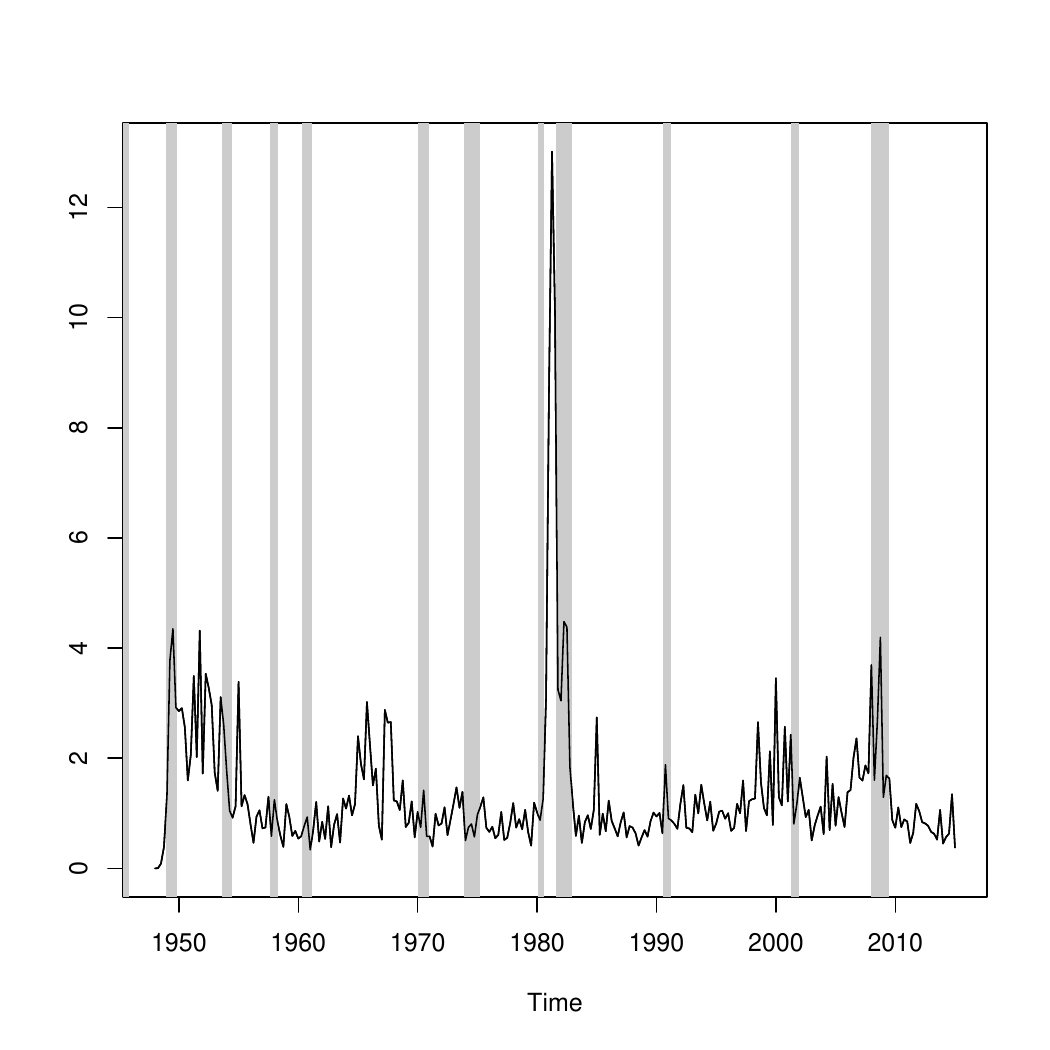}
\caption{Overall}\label{fig:2h}
\end{subfigure}
\caption{Determinant of time-varying variance-covariance matrix of the innovations to the state equation. Values are obtained by taking the mean exponential of the demeaned log-determinant across equations. Gray shaded areas refer to US recessions dated by the NBER business cycle dating committee.}
\label{fig:concovtrace}
\end{figure}

In what follows, we examine the posterior mean  of the determinant of the time-varying variance-covariance matrix of the innovations in the state equation \citep{cogley2005drifts}. For each draw of $\boldsymbol{\Omega}_{it} = \text{diag}(\theta_{i1,t },\dots,\theta_{iK_i,t})$ we compute the logarithm of the determinant and subtract the mean across time. Large values of this measure point towards a pronounced degree of time-variation in the autoregressive coefficients of the corresponding equations.  The results are provided in \autoref{fig:concovtrace} for each equation and the full system.

For all variables we see at least one prominent spike during the sample period indicating a structural break. Most spikes in the determinant occur around 1980, when then Fed chairman Paul Volcker sharply increased short-term interest rates to fight inflation. Other breaks relate to the dot-com bubble in the early 2000s (consumption), the oil price crisis and stock market crash in the early 1970s (hours worked) and another oil price related crisis in the early 1990s. Also, the transition from positive interest rates to the zero lower bound in the midst of the global financial crisis is indicated by a spike in the determinant. That we can relate spikes to historical episodes of financial and economic distress lends further confidence in the modeling approach. Albeit among these periods, the early 1980s seem to have constituted by far the most severe rupture for the US economy, the analysis reveals several further, variable-dependent structural breaks. A model that assumes common dynamics of the coefficients would not be able to pick these up, which emphasizes the flexibility of the proposed approach.

\subsection{Impulse responses to a monetary policy shock}
In this section we examine the dynamic responses of a set of macroeconomic variables to a contractionary monetary policy shock. The monetary policy shock is calibrated as a 100 basis point (bp) increase in the FFR and identified using a Cholesky ordering with the variables appearing in exactly the same order as mentioned above. This ordering is in the spirit of \cite{christiano2005} and has been subsequently used in the literature \citep[see][for an excellent survey]{Coibion2012}.  

We proceed in two stages. First, we show slices of impulse responses for the 4-step, 8-step and 12-step ahead forecast horizon. This allows us to get an overall impression of the time variation in the impulse response functions. In the second stage we zoom in and provide  the full set of impulse responses for two sub-sets of the sample, namely the pre-Volcker period from 1947Q4 to 1979Q1 and the rest of the sample. All impulse response functions are calculated assuming that the shocks to the states are set to their expected value, i.e., zero. Hence we follow \cite{koop2009evolution} and neglect the fact that parameters might be changing over the impulse response horizon. Compared to dynamic forecasts, this simpler strategy is computationally less involved and in light of the fact that our model detects rather few (but large) structural breaks, appears to be reasonable.

\begin{sidewaysfigure}[p]
\includegraphics[width=\textwidth]{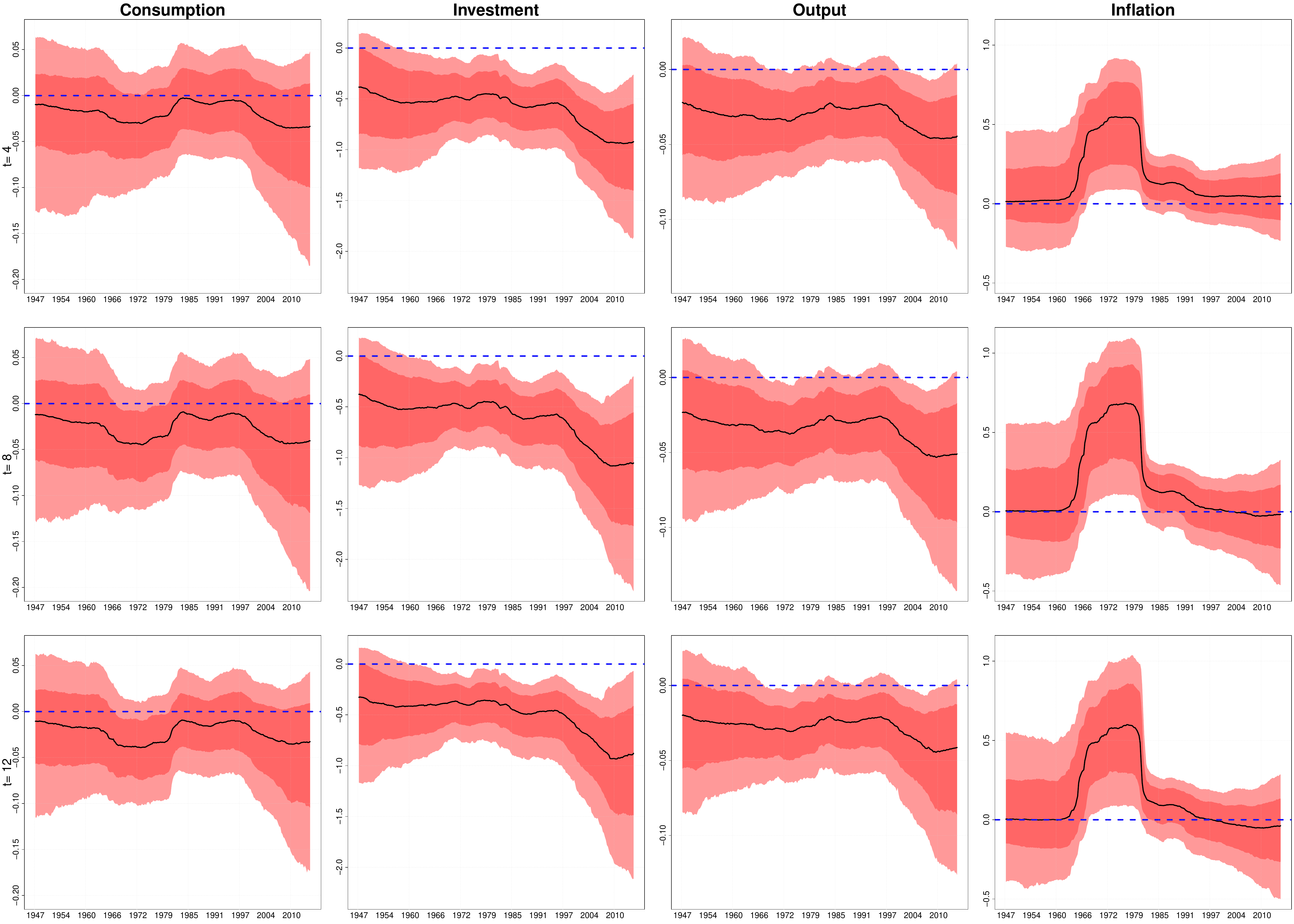}
\caption{Posterior median responses to a $+100$ bp monetary policy shock, after 4 (top panels), 8 (middle panels) and 12 (bottom panels) quarters. Shaded areas correspond to 90\% (dark red) and 68\% (light red) credible sets.}\label{fig:irf1}
\end{sidewaysfigure}


\begin{sidewaysfigure}[p]
\includegraphics[width=\textwidth]{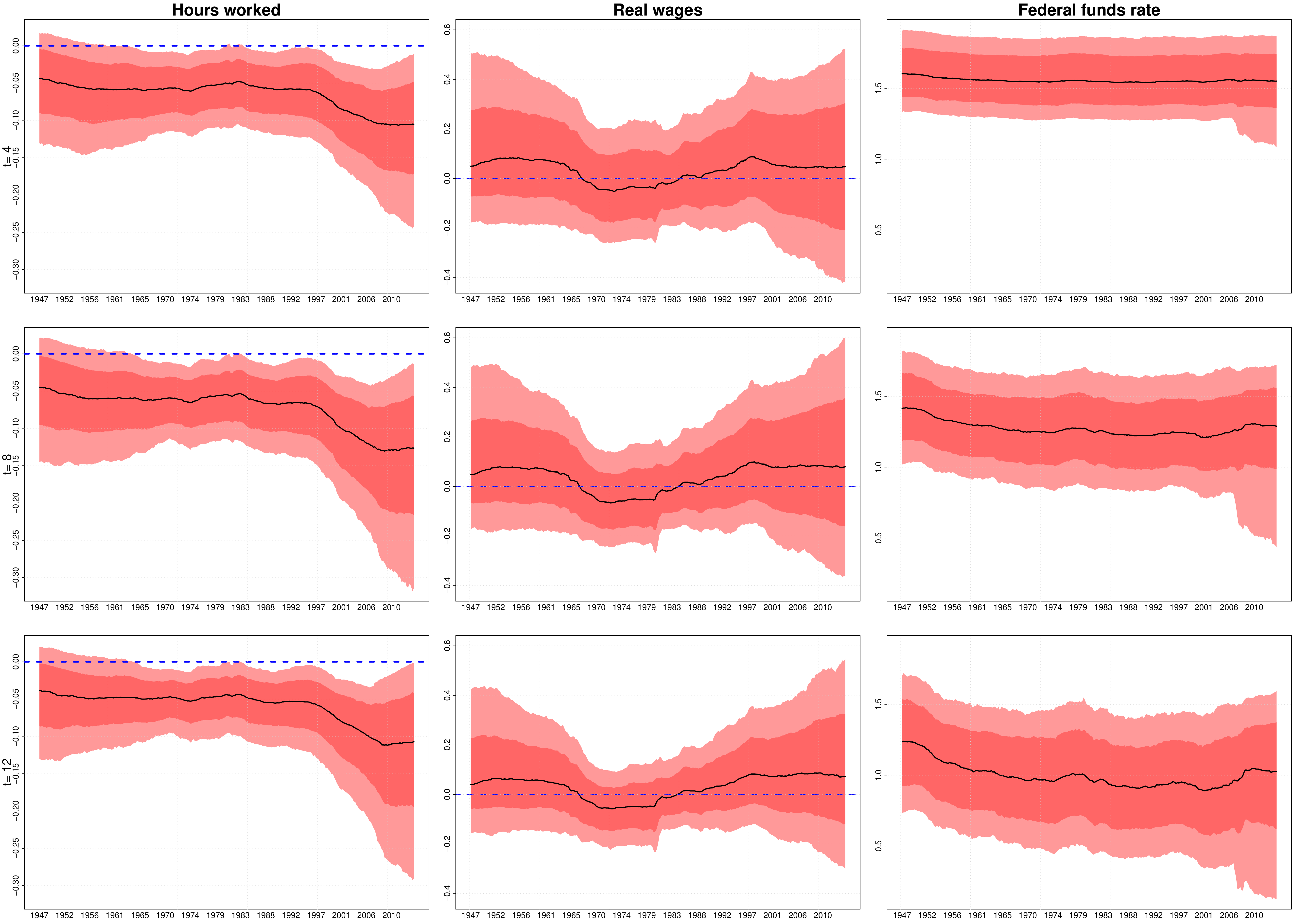}
\caption{Posterior median responses to a $+100$ bp monetary policy shock, after 4 (top panels), 8 (middle panels) and 12 (bottom panels) quarters. Shaded areas correspond to 90\% (dark red) and 68\% (light red) credible sets.}\label{fig:irf2}
\end{sidewaysfigure}

In \autoref{fig:irf1} and \autoref{fig:irf2}, we see the overall effects of a monetary tightening: As investment growth decelerates, hours worked decline and overall output growth decreases. 
 These results are reasonable from an economic perspective. Also, estimated effects on output growth and inflation are comparable to those of \citet{Baumeister2013} who use a TVP-VAR framework and US data. Results on consumption growth and real wages are accompanied by large credible intervals. 
 
Looking at time variation, the results indicate stronger (i.e., more negative) effects of monetary policy for the most recent part of our sample period. More precisely and starting in the late-1990s, effects on  consumption, investment and output growth gradually decrease until the end of our sample period. Most interestingly, though, are responses of inflation: They sharply increase in the late 1960s resulting in a pronounced ``price puzzle'', remain constant in the 1970s and start declining sharply with the onset of the Volcker-period. This pattern is also mirrored in responses of real wages, which are only  negative during the period of pronounced inflation effects, while positive during the rest of the sample period. The results in \autoref{fig:irf1} and \autoref{fig:irf2} thus reveal time variation in the effects of a monetary policy shock and -- more importantly -- that these are variable specific and can be both gradual and abrupt. 
 
\begin{figure}[p]
\begin{minipage}{1\linewidth}~\\
\centering \textbf{1947Q4 to 1979Q1}
\end{minipage}\\[.5em]
\begin{minipage}[b]{0.246\linewidth}
\centering \includegraphics[clip, trim=20 45 20 50, width=\linewidth]{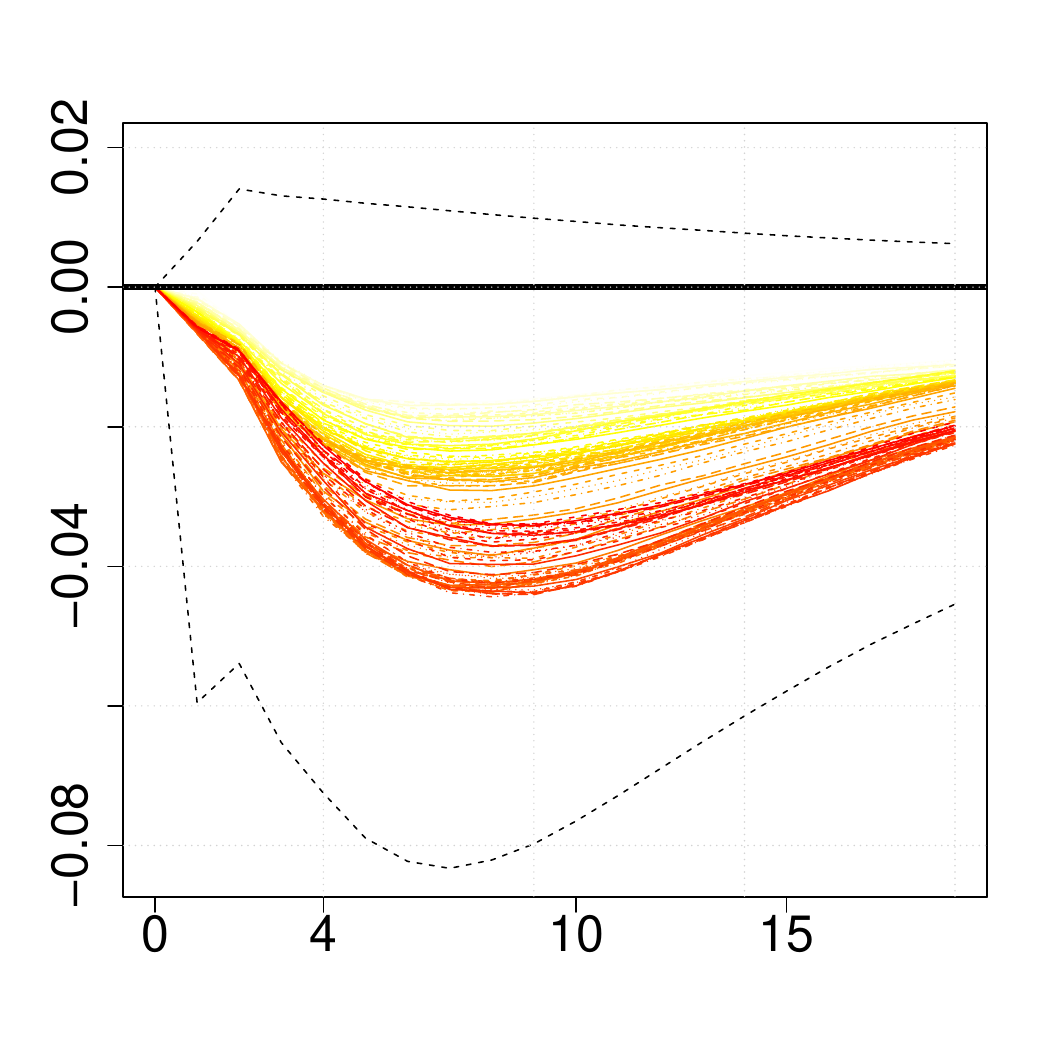}
Consumption
\end{minipage}%
\begin{minipage}[b]{0.246\linewidth}
\centering \includegraphics[clip, trim=20 45 20 50, width=\linewidth]{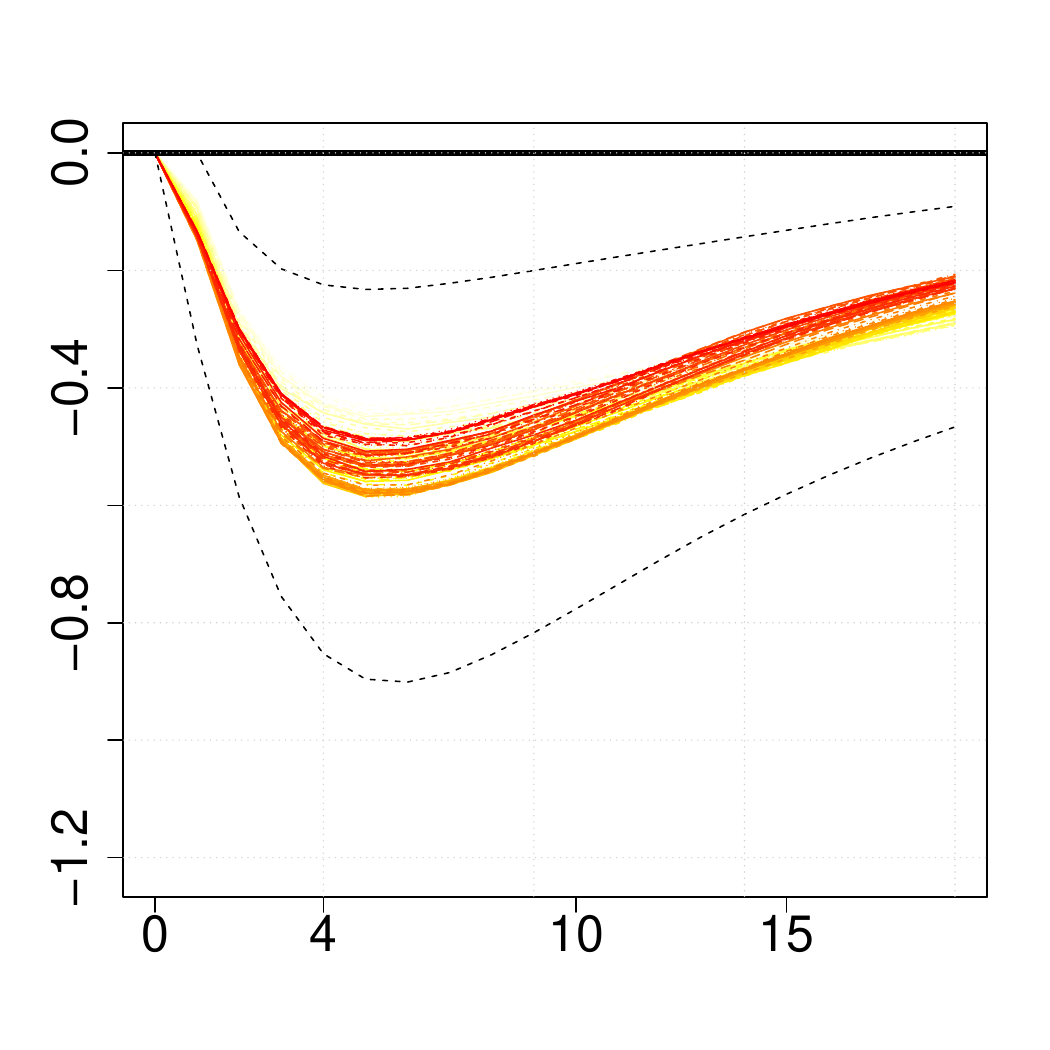}
Investment
\end{minipage}
\begin{minipage}[b]{0.246\linewidth}
\centering \includegraphics[clip, trim=20 45 20 50, width=\linewidth]{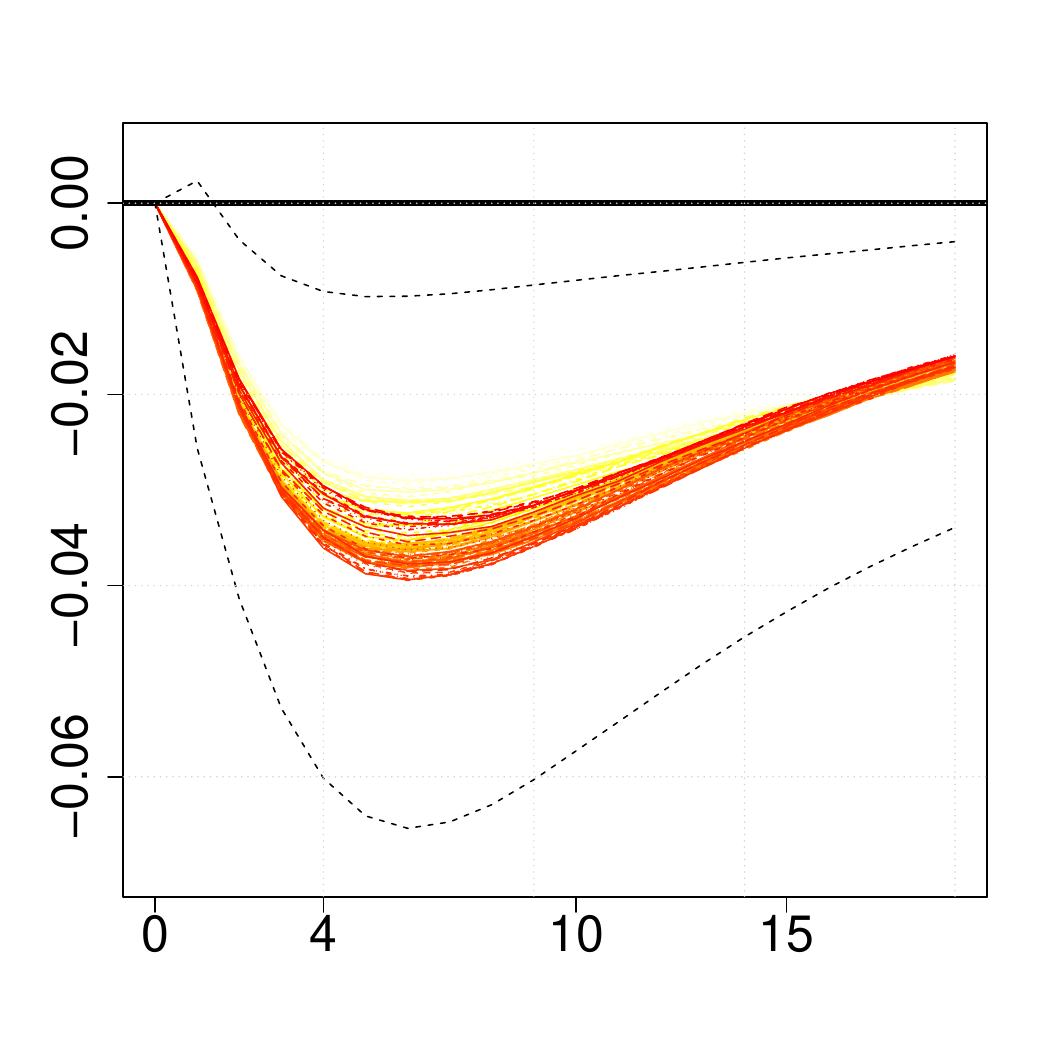}
Output
\end{minipage}
\begin{minipage}[b]{0.246\linewidth}
\centering \includegraphics[clip, trim=20 45 20 50, width=\linewidth]{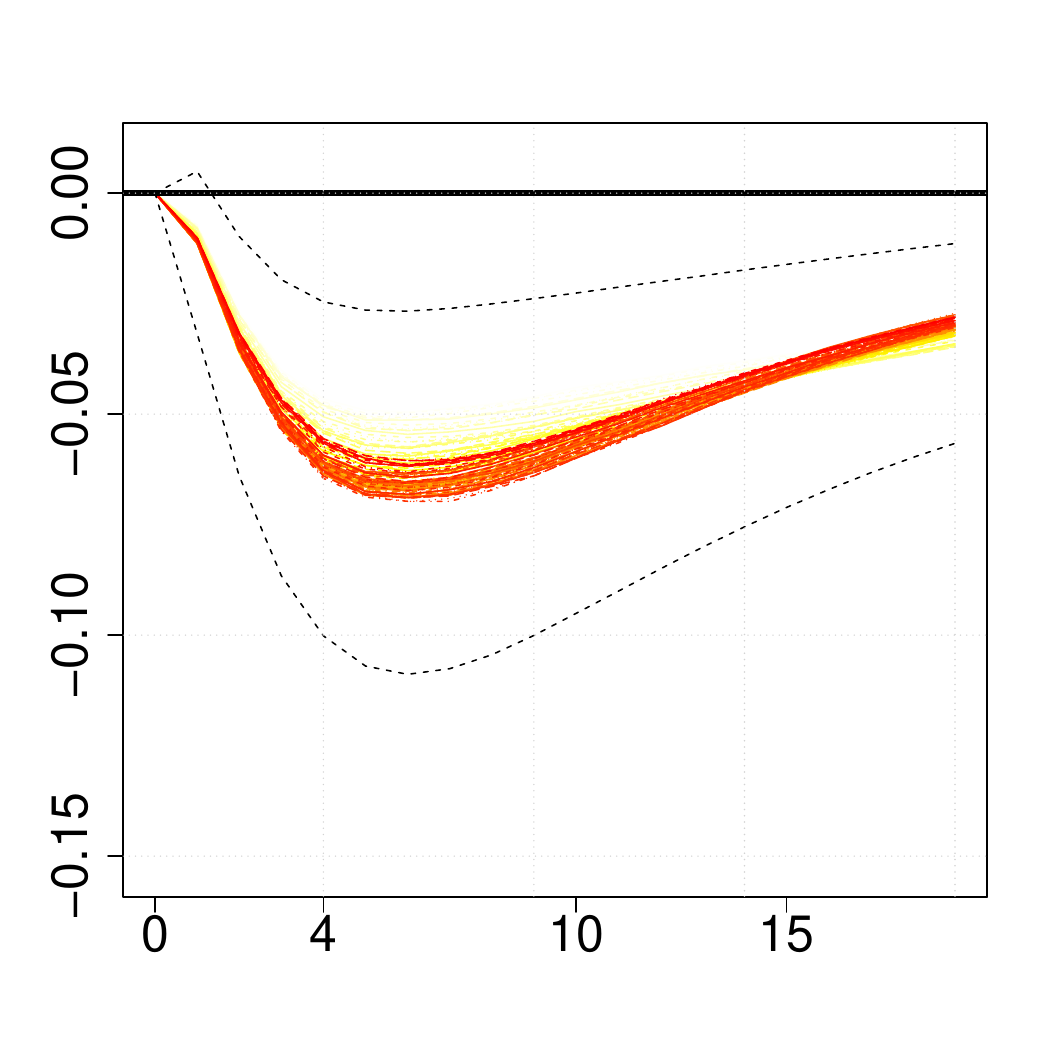}
Hours worked
\end{minipage}
\vspace{-.5em}

\centering
\begin{minipage}[b]{0.246\linewidth}
\centering \includegraphics[clip, trim=20 45 20 50, width=\linewidth]{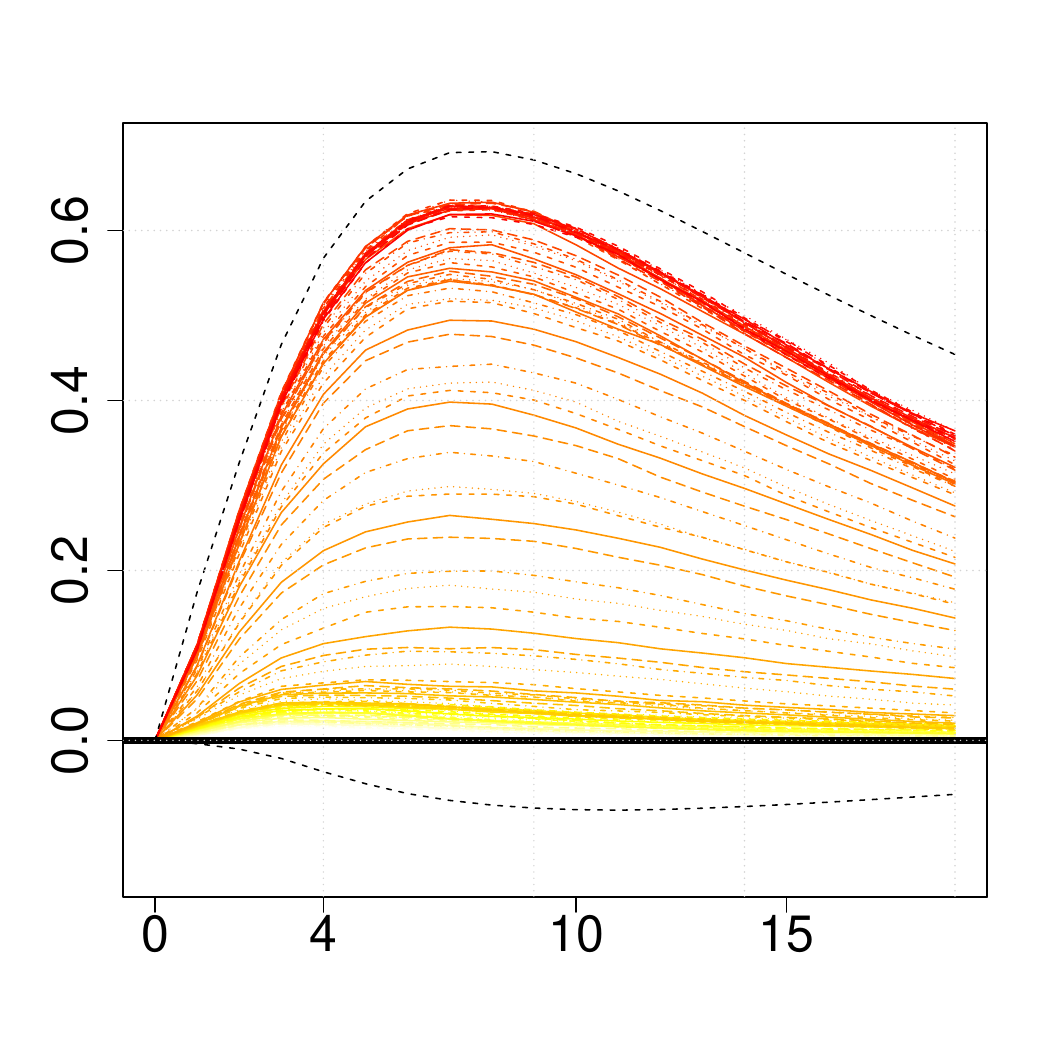}
Inflation
\end{minipage}
\begin{minipage}[b]{0.246\linewidth}
\centering \includegraphics[clip, trim=20 45 20 50, width=\linewidth]{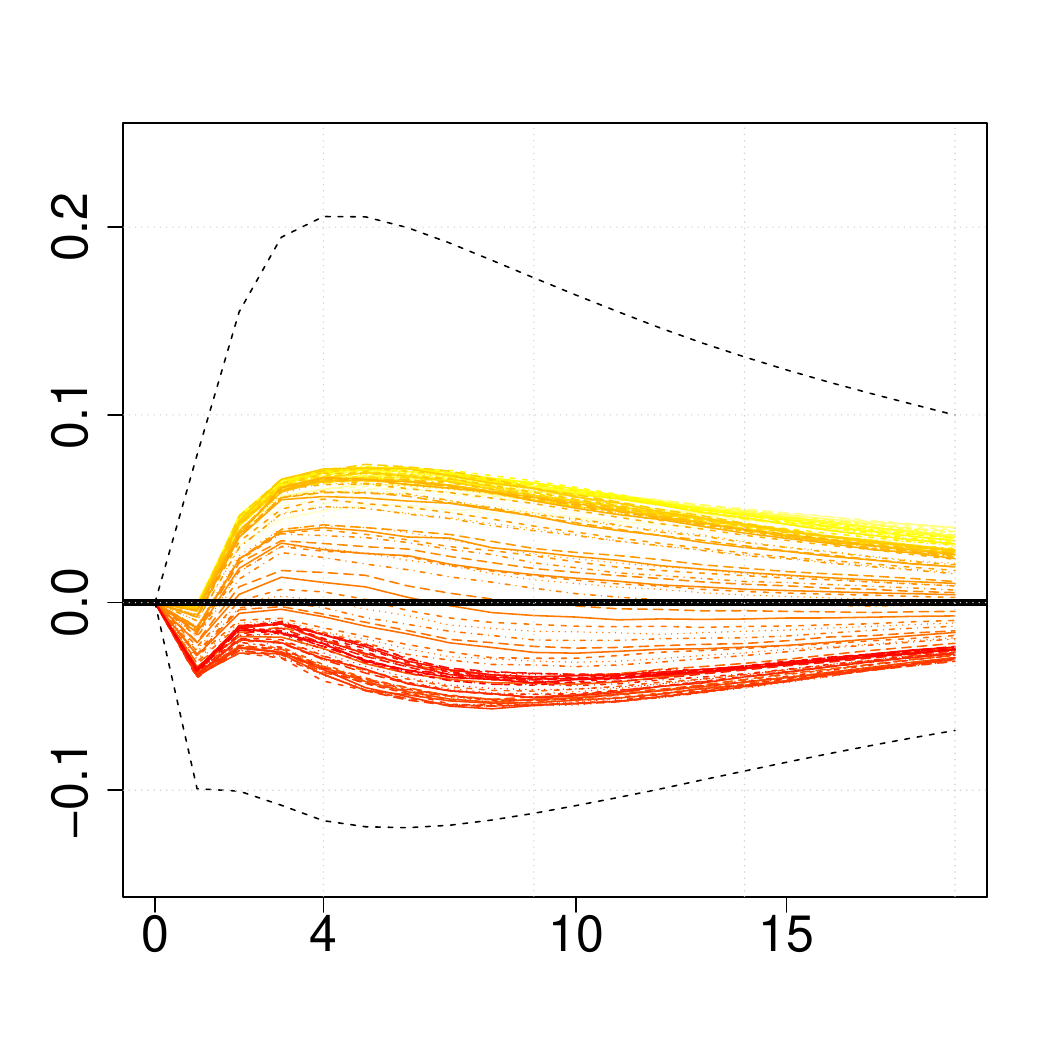}
Real wages
\end{minipage}
\begin{minipage}[b]{0.246\linewidth}
\centering \includegraphics[clip, trim=20 45 20 50, width=\linewidth]{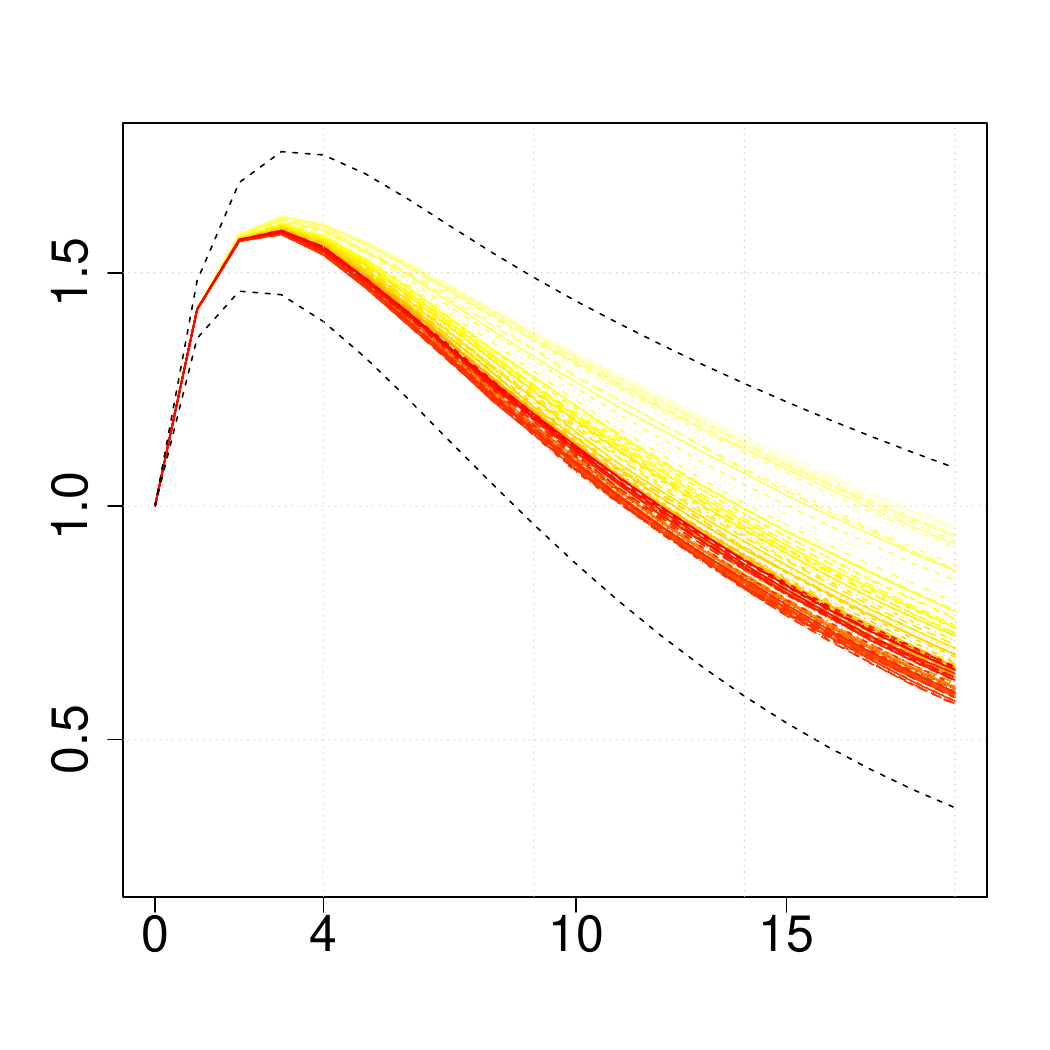}
FFR
\end{minipage}\\

\vspace{2em}

\begin{minipage}{1\linewidth}~\\
\centering \textbf{1979Q2 to 2014Q4}
\end{minipage}\\[.5em]

\begin{minipage}[b]{0.246\linewidth}
\centering \includegraphics[clip, trim=20 45 20 50, width=\linewidth]{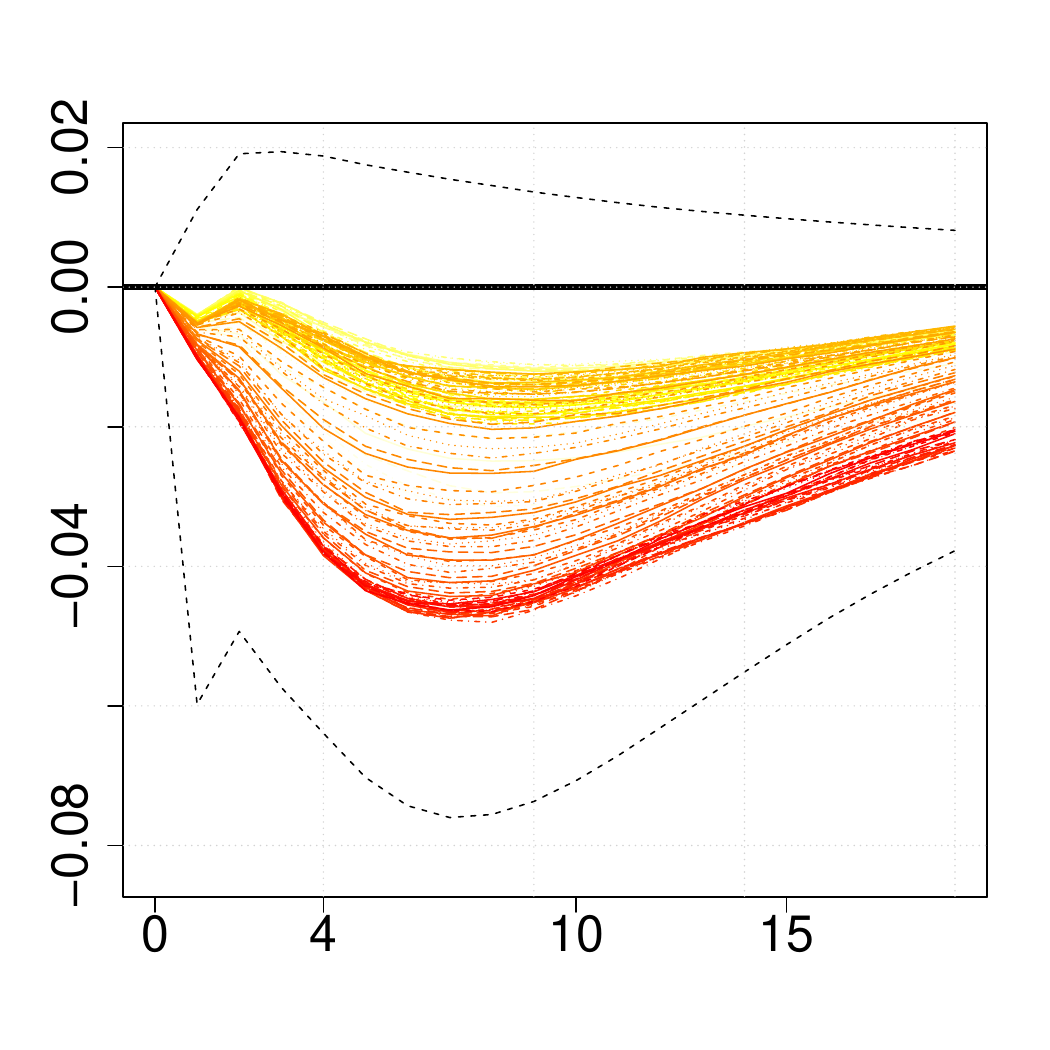}
Consumption
\end{minipage}%
\begin{minipage}[b]{0.246\linewidth}
\centering \includegraphics[clip, trim=20 45 20 50, width=\linewidth]{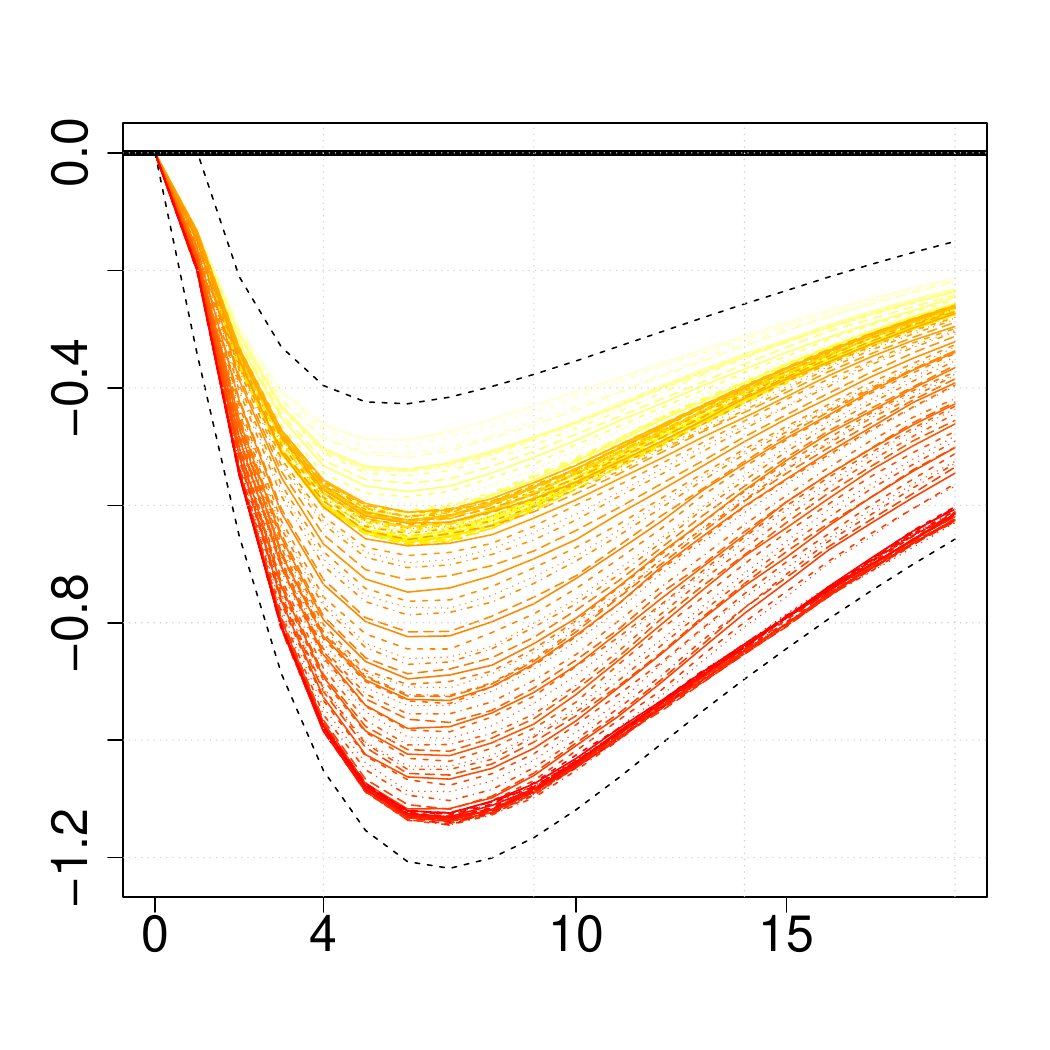}
Investment
\end{minipage}
\begin{minipage}[b]{0.246\linewidth}
\centering \includegraphics[clip, trim=20 45 20 50, width=\linewidth]{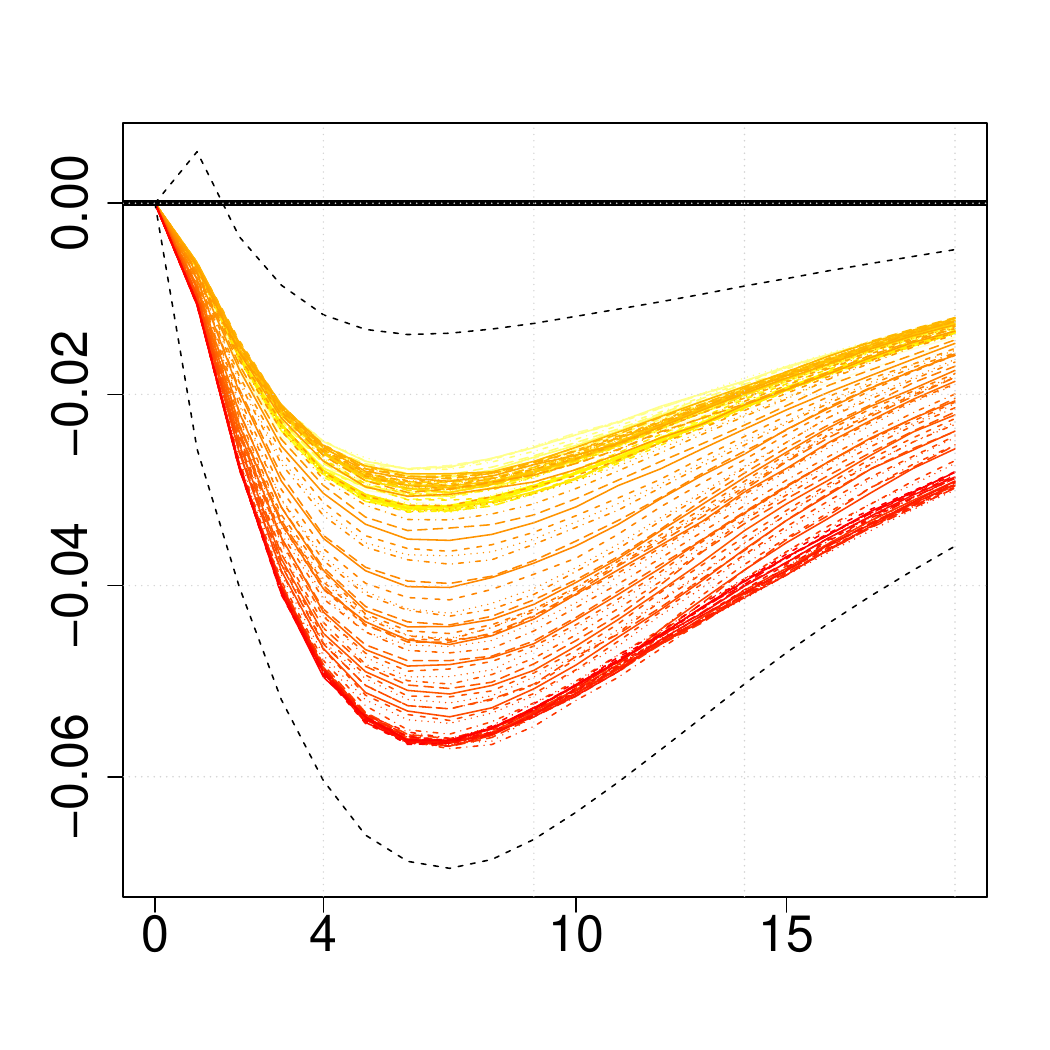}
Output
\end{minipage}
\begin{minipage}[b]{0.246\linewidth}
\centering \includegraphics[clip, trim=20 45 20 50, width=\linewidth]{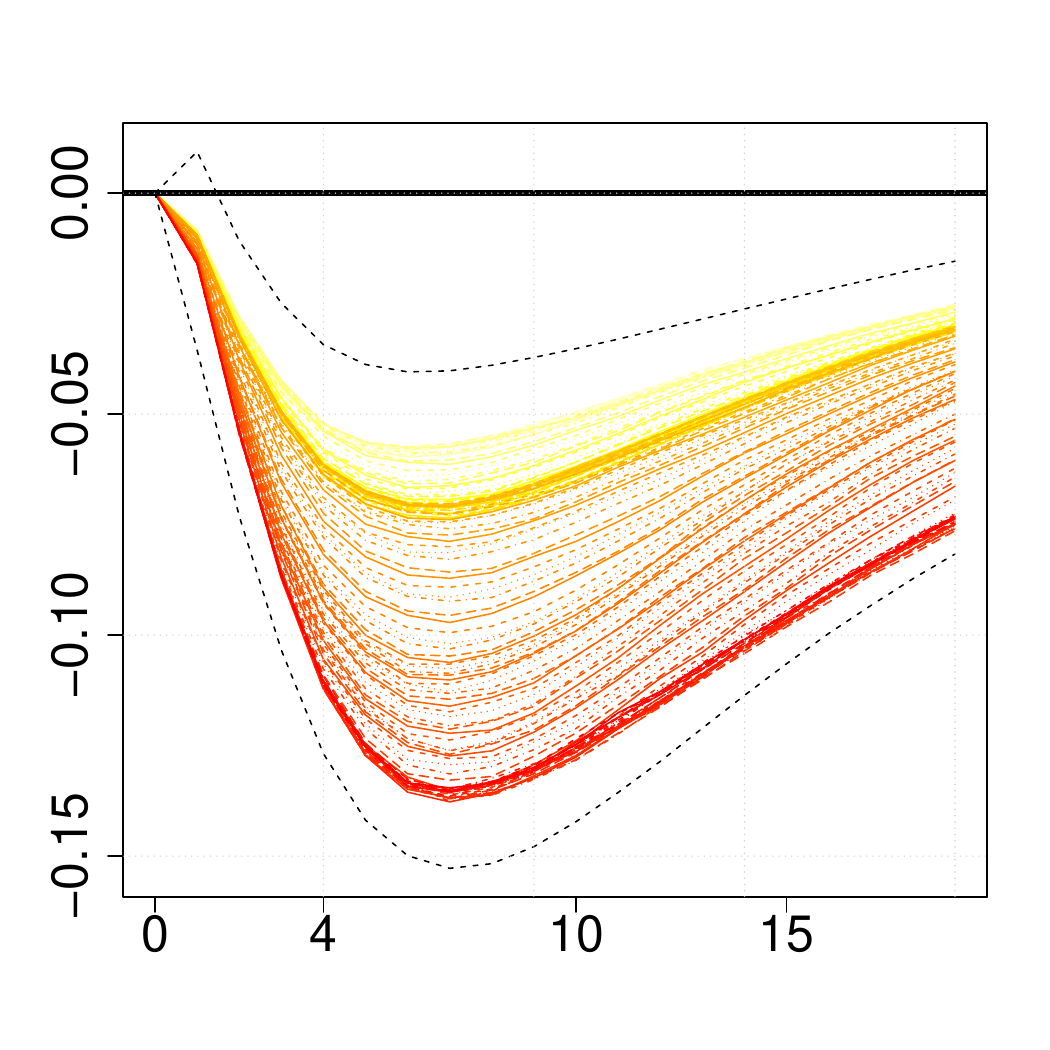}
Hours worked
\end{minipage}

\vspace{.5em}

\centering
\begin{minipage}[b]{0.246\linewidth}
\centering \includegraphics[clip, trim=20 45 20 50, width=\linewidth]{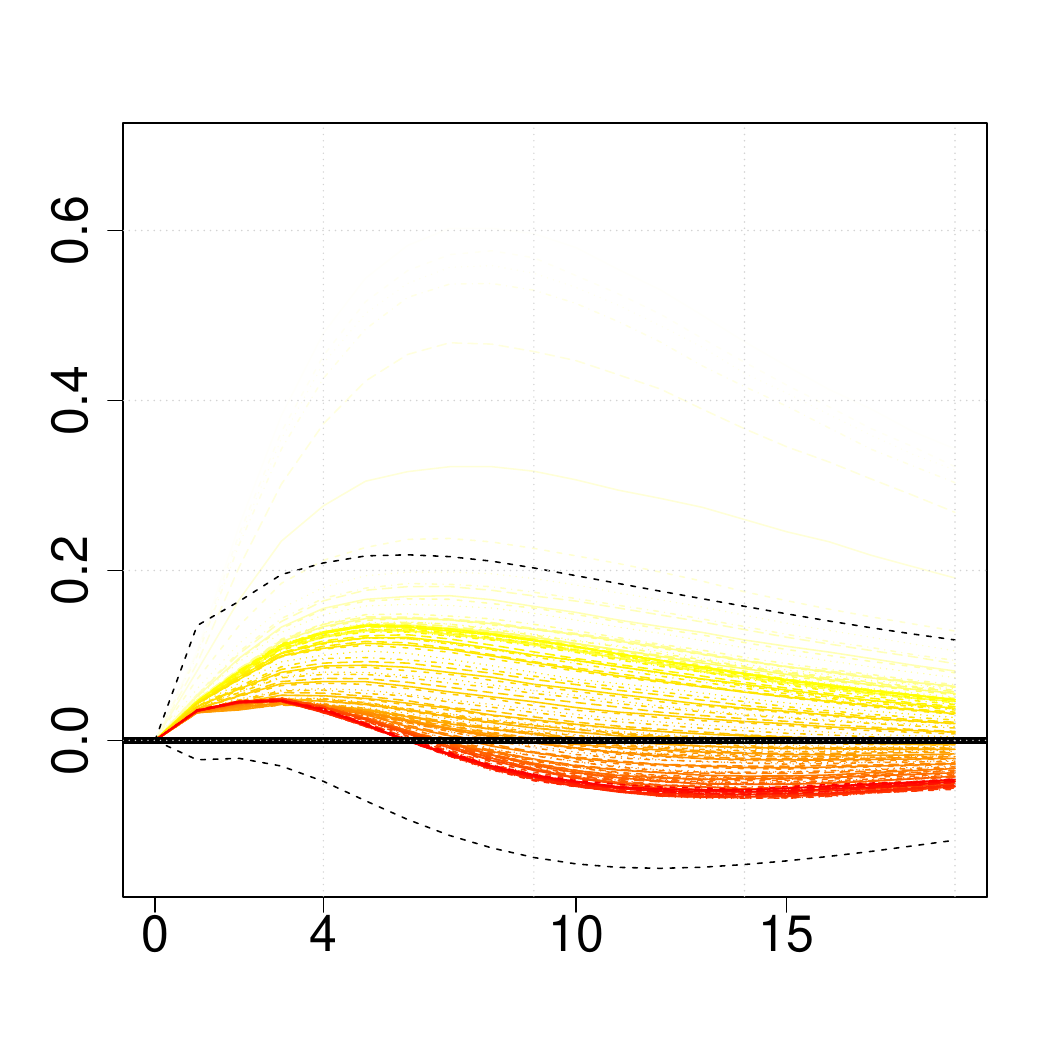}
Inflation
\end{minipage}
\begin{minipage}[b]{0.246\linewidth}
\centering \includegraphics[clip, trim=20 45 20 50, width=\linewidth]{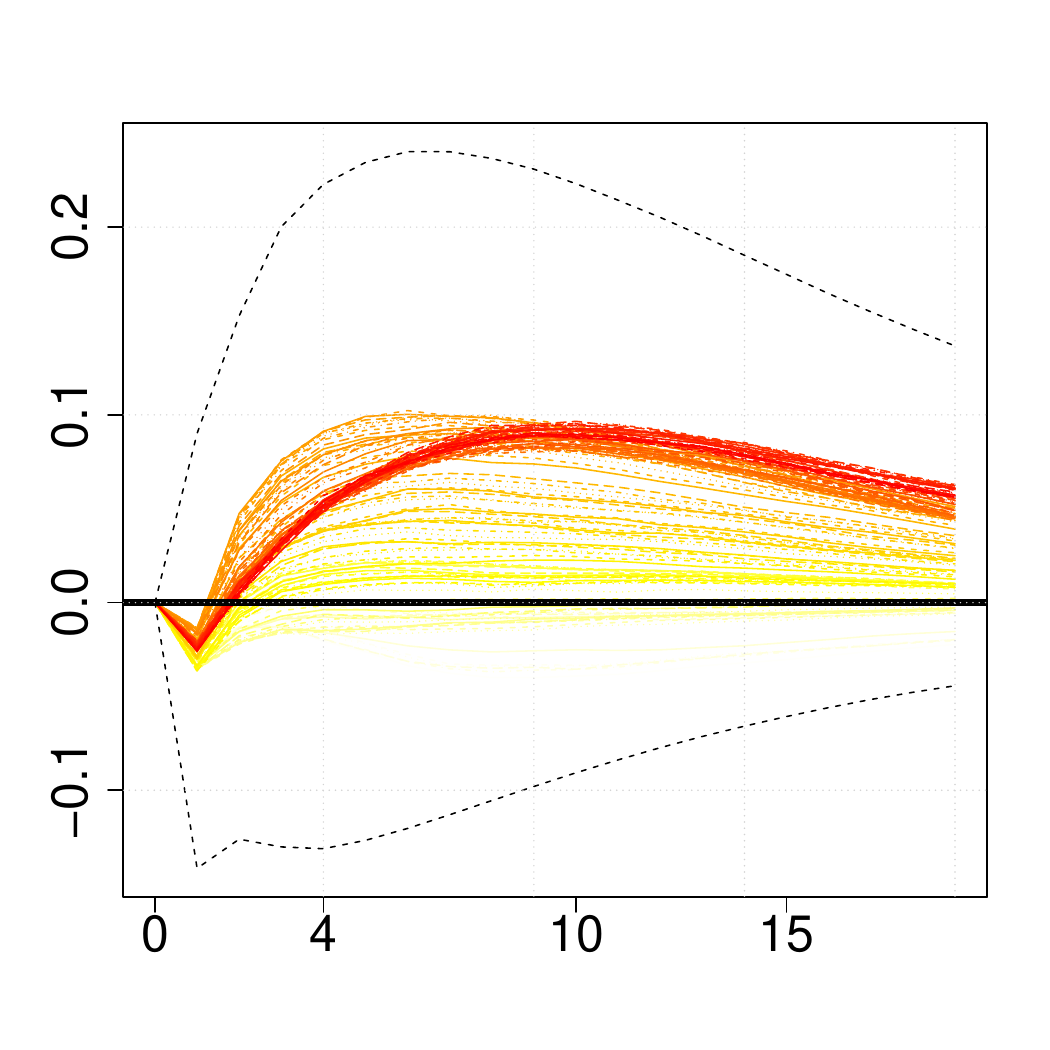}
Real wages
\end{minipage}
\begin{minipage}[b]{0.246\linewidth}
\centering \includegraphics[clip, trim=20 45 20 50, width=\linewidth]{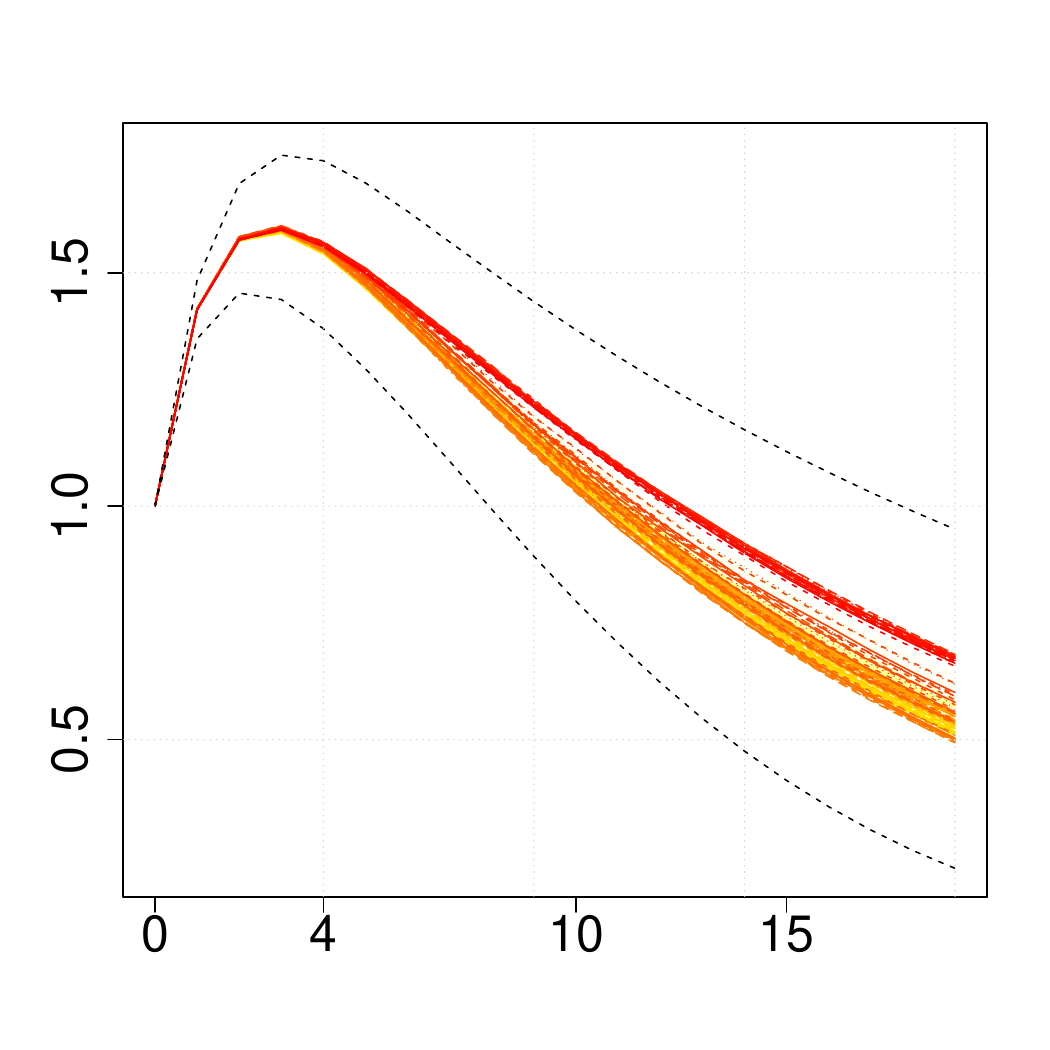}
FFR
\end{minipage}

\caption{Posterior median impulse response functions over two sample splits, namely the pre-Volcker period (1947Q4 to 1979Q1) and the rest of the sample period (1979Q2 to 2014Q4). The coloring of the impulse responses refer to their timing: light yellow stands for the beginning of the sample split, dark red stands for the end of sample split. For reference, 68\% credible intervals over the average of the sample period provided (dotted black lines).}\label{fig:irf_volcker}
\end{figure}

We next zoom in and focus on two sub-sets of the sample, namely the pre-Volcker period from 1947Q4 to 1979Q1 and the rest of the sample.\footnote{The split into two sub-sets is conducted for interpretation purposes only. For estimation, the entire sample has been used.} The time-varying impulse responses -- as functions of horizons -- are displayed in \autoref{fig:irf_volcker}. 
We investigate whether the size and the shape of responses varies between and within the two sub-samples. For that purpose, we show median responses over the first sample split in the top row and for the second part of the sample in the bottom row.
Impulse responses that belong to the beginning of a sample split are depicted in light yellow, those that belong to the end of the sample period in dark red. To fix ideas, if the size of a response increases continuously over time we should see a smooth darkening of the corresponding impulse from light yellow to dark red. 

Considering the first sub-period from 1947Q4 to 1979Q1,  one of the variables that shows a great deal of variation in magnitudes is the response of inflation. Here, effects become increasingly positive the further one moves from 1947Q4 to 1979Q1  and the shades of the responses turn continuously darker.  While overall credible sets for the sub-sample are wide, positive responses for inflation and thus the price puzzle are estimated over the period from the mid-1960s to the beginning of the 1980s (see also \autoref{fig:irf1}). A similar picture arises when looking at consumption growth. During the first sample split, effects become increasingly more negative, but responses are only precisely estimated for the period from the mid-1960s to the beginning of the 1980s. This might be explained by the fact that the monetary policy driven increase in inflation spurs consumption since saving becomes less attractive.

In the bottom panel of \autoref{fig:irf_volcker}, we focus on the results over the more recent second sample split from 1979Q2 to 2014Q4. Paul Volcker's fight against inflation had some bearings on overall macroeconomic dynamics in the USA. With the onset of the 1980s, the aforementioned price puzzle starts to disappear (in the sense that effects are surrounded by wide credible sets and medium responses increasingly negative).  There is also a great deal of time variation evident in other responses  which are mostly becoming increasingly negative. Put differently, the effectiveness of monetary policy seems to be higher in the more recent sample period than before. This can be seen by effects on hours worked, investment growth and output growth. That the effects of a hypothetical monetary policy shock on output growth are particular strong after the crisis corroborates findings of \citet{Baumeister2013} and \citet{Feldkircher2016}. The latter argue that this is related to the zero lower bound period: after a prolonged period of unaltered interest rates, a deviation from the (long-run) interest rate mean can exert considerable effects on the macroeconomy.
 
 
 

\section{Closing remarks}
\label{sec:conclusion} 
This paper puts forth a novel approach to estimating large-scale  time-varying parameter models with mixture innovations in a Bayesian framework. We propose approximating the exact indicators that control which mixture component to use by a threshold process where the threshold variable is the absolute period-on-period change of the corresponding states. This implies that if the (proposed) change is sufficiently large, the corresponding variance is set to a value greater than zero. Otherwise, it is set close to zero which implies that the states remains virtually constant between two points in time. Our framework is capable of discriminating between a plethora of competing specifications, most notably models that feature moderately many, few, or even no structural breaks in the regression parameters.

The merits of our approach are illustrated by two applications. The first application serves as a means to assess the forecasting capabilities of the proposed model while the second model illustrates how the framework can be used to perform structural analysis. In the first application, we show that our model performs well when used to predict the US term structure of interest rates.  Our results indicate that the model yields precise forecasts, especially so during more volatile times such as witnessed in 2008 and during the debt ceiling crisis in 2011. For that period, the forecast gain over simpler models is particularly high. 

For the second application, we turn to US macroeconomic data. We investigate whether reduced-form parameters vary over time by considering the time-varying determinant of the posterior variance-covariance matrix of the state innovations. This analysis suggests several variable specific structural breaks in the reduced form relationships, with the Volcker period marking the most severe rupture for the US economy. Examining the effects of a contractionary monetary policy shock, we see considerable time variation in the impulse responses. Our results indicate abrupt changes of effects on inflation. More specifically, we find significant evidence for a severe price puzzle during episodes of the pre-Volcker period, whereas  the puzzle disappears in the second half of our sample. Effects on other variables such as output and investment growth as well as hours worked change more gradually, reaching a trough during  the period after the global financial crisis. For that period, a hypothetical deviation from the zero lower bound would create  pronounced effects on the wider macroeconomy. These findings highlight the importance to account for different dynamics of the underlying variables in order to adequately capture the complex interaction of the macroeconomy -- a salient feature of our modeling framework.


\section{Acknowledgments}
We sincerely thank several anonymous reviewers, the participants of the 7th and 8th European Seminar on Bayesian Econometrics (ESOBE 2016 and 2017), the WU Brown Bag Seminar of the Institute of Statistics and Mathematics, the 3rd Vienna Workshop on High-Dimensional Time Series in Macroeconomics and Finance 2017, the NBP Workshop on Forecasting 2017, and in particular Sylvia Fr\"uhwirth-Schnatter, Hedibert Freitas Lopes, Herman van Dijk, and Helga Wagner for many helpful comments and suggestions that improved the paper significantly.

\singlespacing
\bibliographystyle{./bibtex/econometrica}
\bibliography{./bibtex/tvp_var}
\addcontentsline{toc}{section}{References}

\onehalfspacing
\begin{appendices}
\section{Convergence and mixing properties}\label{sec: convergence}
Here, we assess convergence of our proposed algorithm for the US macroeconomic dataset. As mentioned in the main part of the paper, convergence characteristics closely resemble those typically reported when standard TVP-VARs with SV are used. To assess mixing and convergence properties of the thresholds, \autoref{tab:convergence_2} shows the empirical distribution of inefficiency factors and the \cite{raftery1992many} diagnostic of the total number of runs required to achieve a certain level of precision. The parameters of the diagnostic are specified as in \cite{primiceri2005time}.\footnote{The quantiles are set equal to 0.025, the desired degree of accuracy is 0.025, and the probability of achieving the required accuracy is 0.95.}

The table indicates that inefficiency factors across equations appear to be favorable (i.e., well below 50) for all covariates. Notice that the marginal posteriors of selected thresholds feature estimated inefficiency factors of one, indicating virtually no autocorrelation. Considering the required number of runs to achieve a certain level of precision reveals that this is far below the actual number of iterations in practically all cases.

\begin{table}[h]
\caption{Empirical distribution across covariates within an equation of selected convergence metrics for the thresholds: US macroeconomic data.}\label{tab:convergence_2}
\scalebox{0.89}{
\centering
\begin{tabular}{lrrrrrrrrrr}
  \toprule
    &\multicolumn{5}{c}{Inefficiency factors}&&\multicolumn{4}{l}{Required number of runs}\\

 & Low$_{10}$ & Median & High$_{90}$ & Min & Max & Low$_{10}$ & Median & High$_{90}$ & Min & Max \\ 
  \hline
consumption & 1 & 1 & 5 & 1 & 7 & 928 & 1375 & 2913 & 907 & 3945 \\ 
  investment & 1 & 7 & 13 & 1 & 31 & 2222 & 3112 & 4725 & 1162 & 4746 \\ 
  output & 1 & 1 & 1 & 1 & 2 & 922 & 968 & 2788 & 907 & 6360 \\ 
  hours & 1 & 1 & 2 & 1 & 7 & 928 & 2129 & 3538 & 907 & 5064 \\ 
  inflation & 1 & 1 & 14 & 1 & 16 & 1094 & 2270 & 4319 & 922 & 4780 \\ 
  real.wage & 1 & 9 & 19 & 1 & 49 & 1415 & 2825 & 4606 & 1242 & 5415 \\ 
  interest.rate & 1 & 1 & 10 & 1 & 19 & 922 & 1242 & 2896 & 922 & 3390 \\ 
   \hline
\end{tabular}
}
\end{table}
\end{appendices}

\end{document}